\documentclass[reprint, showpacs]{revtex4-1}
\usepackage{graphicx}
\usepackage{dcolumn}
\usepackage{bm}
\usepackage{amssymb}
\usepackage{amsmath}
\usepackage{tabularx}
\usepackage{hyperref} 

\usepackage{xcolor}

\usepackage{natbib}
\usepackage{placeins}

\newcommand{\be}{\begin{equation}}
\newcommand{\ee}{\end{equation}}
\newcommand{\bea}{\begin{eqnarray}}
\newcommand{\eea}{\end{eqnarray}}

\begin{document}
\title{Phenomenology of quantum eigenstates in mixed-type systems: lemon billiards with complex phase space structure}

\author{\v Crt Lozej$^{1,2}$}
\author{Dragan Lukman$^1$}
\author{Marko Robnik$^1$}

\affiliation{$^1$ CAMTP - Center for Applied Mathematics and Theoretical
  Physics, University of Maribor, Mladinska 3, SI-2000 Maribor, Slovenia\\
  $^2$ Max Planck Institute for the Physics of Complex systems,
  N\"othnitzer Str. 38, D-01187 Dresden, Germany}


\date{\today}

\begin{abstract} The boundary of the lemon billiards is defined by
  the intersection of two circles of equal unit radius with the
  distance $2B$ between their centers, as introduced by
  Heller and Tomsovic in Phys. Today {\bf 46} 38 (1993). 
  This paper is a continuation of our recent papers
  on classical and quantum ergodic lemon billiard ($B=0.5$) with strong
  stickiness effects (Phys. Rev. E {\bf 103} 012204 (2021)),
  as well as on the three billiards with a simple
  mixed-type phase space and no stickiness (Nonlinear Phenomena
  in Complex Systems {\bf 24} No 1,p 1-18 (2021)).
  Here we study two classical and quantum lemon billiards,
  for the cases $B=0.1953,\; 0.083$, which are mixed-type billiards
  with a complex structure of phase space, without significant
  stickiness regions. A preliminary study of their spectra
  was published recently (Physics {\bf 1} 1-14 (2021)).
  We calculate a very large number ($10^6$)
  of consecutive eigenstates and their Poincar\'e-Husimi
  (PH) functions, and analyze their localization properties by
  studying the entropy localization measure and the normalized
  inverse participation ratio. We introduce an
  overlap index, which measures the degree of the overlap of
  PH functions with classically regular and chaotic regions.
  We observe the existence of regular states associated with
  invariant tori and chaotic states associated with the
  classically chaotic regions, and also the mixed-type
  states. We show that in accordance with the Berry-Robnik picture
  and the principle of uniform semiclassical condensation
  of PH functions the relative fraction of mixed-type states
  decreases as a power law with increasing energy, thus
  in the strict semiclassical limit leaving only
  purely regular and chaotic states. Our
  approach offers a general phenomenological overview
  of the structural and localization properties of PH
  functions in quantum mixed-type Hamiltonian systems.
  \end{abstract}


\maketitle

\section{Introduction}
\label{sec: introduction}

Classical generic Hamiltonian systems exhibit both regular and chaotic motion \citep{LichLiebBook}, depending on the initial condition. They are referred to as systems with divided phase space or mixed-type systems, because the phase space is divided into regular and chaotic invariant components, with an intricate hierarchical structure of islands of stability embedded in the chaotic sea. Accordingly, the chaotic sea(s) and islands of stability, comprised of invariant tori, may be combined into disjoint measurable subsets with a positive a Liuoville measure (phase space volume).  According to the correspondence principle, one expects that the eigenstates of the equivalent quantized system should behave similarly. The states may be separated into subsets that correspond to either the chaotic or regular classical dynamics, with a spectral density that is equal to the classical Liouville measure of the corresponding invariant component in the phase space.  The idea was first conjectured by Percival \cite{percival1973}, further elaborated by Berry \cite{berry1977,berry1977semi}  and later developed into the principle of uniform semiclassical condensation (PUSC), see \cite{Rob1998} and references therein. The states may be separated, depending on the overlap with either the regular or chaotic part of the classical phase space, by means of Wigner functions or Husimi functions. Following PUSC, the high-lying eigenstates are supported either on the chaotic sea or the invariant tori forming the islands of stability in the ultimate semiclassical limit. The partial spectrum of the regular states follows Poissonian statistics, while the spectral statistics of the chaotic states are well described by random matrix theory (RMT) \cite{Stoe, Haake}. The whole spectrum may be collectively described by the Berry-Robnik spectral statistics \cite{BerRob1984}. An abundance of numerical evidence corroborates the Berry-Robnik picture and PUCS as its foundation \cite{prosen1993survey,prosen1994numerical,prosen1994semiclassical,li1994statistical,li1995geometry,li1995separating,prosen1995quantum,prosen1999intermediate,veble1999,BatRob2010}. However, a true separation to regular and chaotic states may only be expected in the asymptotic semi-classical limit (where the action is large compared to $\hbar$). Before reaching this asymptotic regime many states will exhibit a mixed behavior, with various tunneling processes between the structures of the classical phase space  and much less is known about this regime despite its rich and interesting phenomenology. For instance, the mixed eigenstates support important physical phenomena like chaos-assisted tunneling \cite{tomsovic1994tunneling}, that has recently been shown to have useful applications in quantum simulation \cite{martinez2021tunneling}. For a general introduction and a rather complete account of quantum chaos we refer to the books by St\"ockmann \cite{Stoe} and  Haake \cite{Haake}, and to the recent review papers on the stationary quantum chaos in generic (mixed-type) systems \cite{Rob2016,Rob2020}.

In this paper, we compute and study the properties of a large number (approximately $10^6$) eigenstates of two examples of lemon billiards with complex divided phase spaces. Billiards are excellent examples of generic model Hamiltonian systems, widely used for studies in quantum chaos.  The lemon billiards were introduced by Heller and Tomsovic \cite{HelTom1993} and have been extensively
studied (including some generalizations)  \cite{Lopac1999,Lopac2001,MakHarAiz2001,ChMoZhZh2013,BunZhZh2015,BCPV2019,Lozej2020,makino2022} 
in the context of classical regular and chaotic dynamics, and as quantum billiards, including our recent works \cite{LLR2021,LLR2021A,LLR2021B}.

\begin{figure}
 \begin{centering}
   \includegraphics[width=9cm]{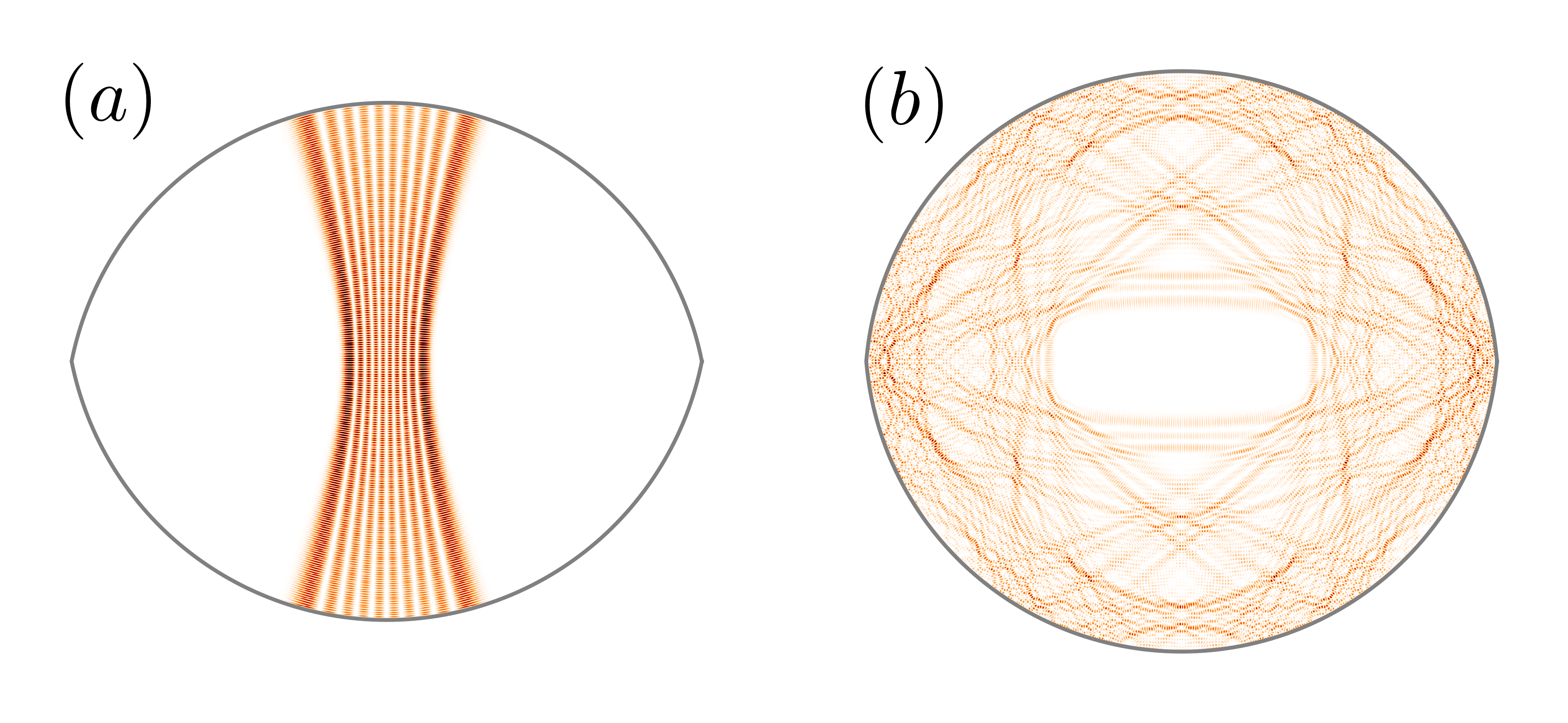}  
   \par\end{centering}
 \caption{Illustration of the two lemon billiards considered in this work together with the probability distributions of a high-lying quantum eigenstate. (a) $B=0.1953$, regular eigenstate at $k=462.122$, (b) $B=0.083$, chaotic eigenstate at $k=457.533$ }
\label{fig: billiards}
\end{figure}

The lemon billiard table is defined as the intersection of two circles of equal unit radius, with a distance of $2B$ between their centers (the construction is explained in Appendix \ref{AppendixA}). In this work, we study classical and quantum mechanics of two lemon billiards, namely $B=0.1953$ and $B=0.083$. They are illustrated in Fig. \ref{fig: billiards}. It must be emphasized that although the lemon billiards belong all to the same
family as for the mathematical definition, individually they have quite different and very rich dynamical properties, which makes them important in both the classical and quantum context. 
The specific parameters were chosen with the following considerations.  The phase space of both billiards consists of one significant chaotic component and several islands of stability. In the $B=0.1953$ case, three major island chains are present, while $B=0.083$ shows a complex web of many island chains. In both cases, stickiness effects (see Refs. \citep{contopoulos2010a,bunimovich2012} for an introduction to the phenomenon) are negligible. This fact is of great significance, since classical stickiness has considerable effects also on the quantum dynamics and structure of the eigenstates as we have recently shown for the case of the presumably ergodic $B=0.5$ billiard in Ref. \cite{LLR2021}. Previously, we have also studied the aspects of quantum chaos in three mixed-type lemon billiards $B=0.42,\; 0.55,\; 0.6$ with a simple structure (only one dominant chaotic component without stickiness regions, coexisting with only one large regular component) in Ref. \cite{LLR2021A}. The cases considered in the present paper allow for an increased complexity and richness of the mixed eigenstates and tunneling effects, but still exclude the effects of strong partial transport barriers, that would classically result in stickiness.
The discovery of these dynamically  different and interesting lemon
billiards has only been made possible thanks to the recent extensive analysis
of Lozej \cite{Lozej2020}. The entire family of classical lemon billiards for
a dense set of about 4000 values of  $B\in [0.01,0.99975]$
(in steps of $dB=0.00025$) has been systematically analyzed as for their
phase space structure and stickiness effects.

A study of quantum energy spectra of the billiards considered in the current work has been recently published in \cite{LLR2021B},
  where specifically we studied the fluctuation of the number of energy levels
  around the mean value determined by the Weyl rule with the perimeter
  corrections, and the energy level spacing distributions for
  all (four) symmetry classes. 

 The main purpose of the present paper is the phenomenological analysis of the eigenstates of the two selected quantum lemon billiards $B=0.1953$ and $0.083$, with the following goals:
(i) To calculate the {\em Poincar\'e-Husimi} (PH) \cite{TV1995, Baecker2004} functions of the eigenstates,
analyze their structure in the phase space in relationship
with the classical phase portrait, and examine the quantum localization
of chaotic eigenstates in the phase space.
(ii) To establish the relationships between various localization and classical-quantum overlap measures in order to present a complete overview of the eigenstates in the PH representation.
(iii) To observe the condensation of the eigenstates on classical invariant components, with progressive energy, in the context of the Berry-Robnik picture of quantum chaos in mixed-type systems \cite{BerRob1984} and the principle of uniform semiclassical condensation \cite{Rob1998}.

The main results are the following. The great majority of PH functions are found to be well-supported either on invariant tori in the regular islands, or on the chaotic component, thus
obeying the principle of uniform semiclassical condensation of
Wigner functions \cite{Rob1998}. PH functions of mixed type exist, and show a wide variety of tunneling states between different classical structures. The distributions of localization and overlap measures may be used to identify the various interesting regimes and quantify their prevalence. The proportion of mixed-type states shows a power law decay with increasing energy in both billiards with an overall exponent of $\gamma \approx 0.3$ and a local variation from 0.1 to 0.5 pertaining to different mixed-state regimes. 

The paper is organized as follows. In Sec. \ref{sec2} we examine the classical dynamical properties of the two lemon billiards under consideration. 
In Sec. \ref{sec3} we define the quantum billiard problem, discuss its numerical solution, and introduce the Poincar\'e - Husimi functions
of the eigenstates.
In Sec. \ref{sec4} we introduce the overlap index $M$ and propose how to
separate the regular and chaotic eigenstates in the sense
of Berry-Robnik \cite{BerRob1984}.  In Sec. \ref{sec5} we introduce and
compute the localization measures of the regular,
chaotic and mixed-type eigenstates and study the structure of their probability distributions and relate it to the Poincar\'e - Husimi functions of individual eigenstates. In \ref{sec6} 
we study the connection between the overlap index $M$ and the
localization measures. In \ref{sec7} we analyze the
energy dependence of the whole picture.
In Sec. \ref{sec8} we summarize and discuss the results and present the
conclusions. Appendix \ref{AppendixA} gives a short overview of the construction and geometry of the lemon billiards and Appendix \ref{AppendixB} presents some relevant results on stickiness and recurrence time statistics.

\section{Phase space structure of lemon billiards}
\label{sec2}

A billiard is a dynamical system which consists of a free
moving point particle confined inside a closed domain $\mathcal{B}$ in Euclidean
space referred to as the billiard table. The particle moves freely inside the billiard table in straight lines
and is specularly reflected when hitting the edge of the table. The family of lemon billiards is formed
by the intersection of two circles of equal unit radius with a
distance of $2B$ between their centers, where $B\in (0,1)$. As usual, we consider the billiard as a discrete dynamical system, taking the boundary as the surface of section. We use the canonical variables to specify the location $q$ and the momentum component $p$ on the boundary at the collision point,
so that the classical phase space is a cylinder $(q,p) \in [0,L]\times(-1,1)$, where $q$ is the arclength (periodic with a period equal to the circumference of the boundary $L$) and $p=\sin(\alpha)$ is the sine of the angle of reflection. For more details on construction of the lemon billiards, the canonical variables, and geometric properties, see Appendix \ref{AppendixA}.
The bounce map, mapping from collision to the next collision,
$(q,p) \rightarrow (q',p')$ is
area preserving as in all billiard systems \cite{Berry1981}.

Due to the two kinks (corners at $y=0$), the Lazutkin invariant tori (related to the boundary glancing orbits) are broken. The period-2 orbit connecting the centers of the two circular arcs is always stable (and therefore surrounded by a regular island) except for the case $B=1/2$, where it is a marginally unstable orbit. This case is presumably ergodic and has been treated in our recent paper \cite{LLR2021}. In all other cases the phase space is divided (mixed-type) with one dominant chaotic sea (that is significantly larger than all other chaotic components) and typically a multitude of islands of stability. Keeping in mind the Berry-Robnik picture, it is useful to compute the relative measure pertaining to chaotic and regular components in the phase space. We compute the relative areas $\chi_c$ and $\chi_r=1-\chi_c$ on the surface of section (two-dimensional phase space of canonical coordinates $(q,p)$) by using the methods presented in Ref. \cite{LozRob2018B}. However, the spectral density of the regular/chaotic states is given by the Liouville measure (volume in the energy surface of the full four-dimensional phase space in Cartesian coordinates $(x,p_x,y,p_y)$). To convert the area into the volume, $\chi_c \rightarrow \rho_c$ we use a formula due to Meyer \cite{meyer1986} given in Appendix \ref{AppendixA}. 

\begin{figure}
 \begin{centering}
   \includegraphics[width=9cm]{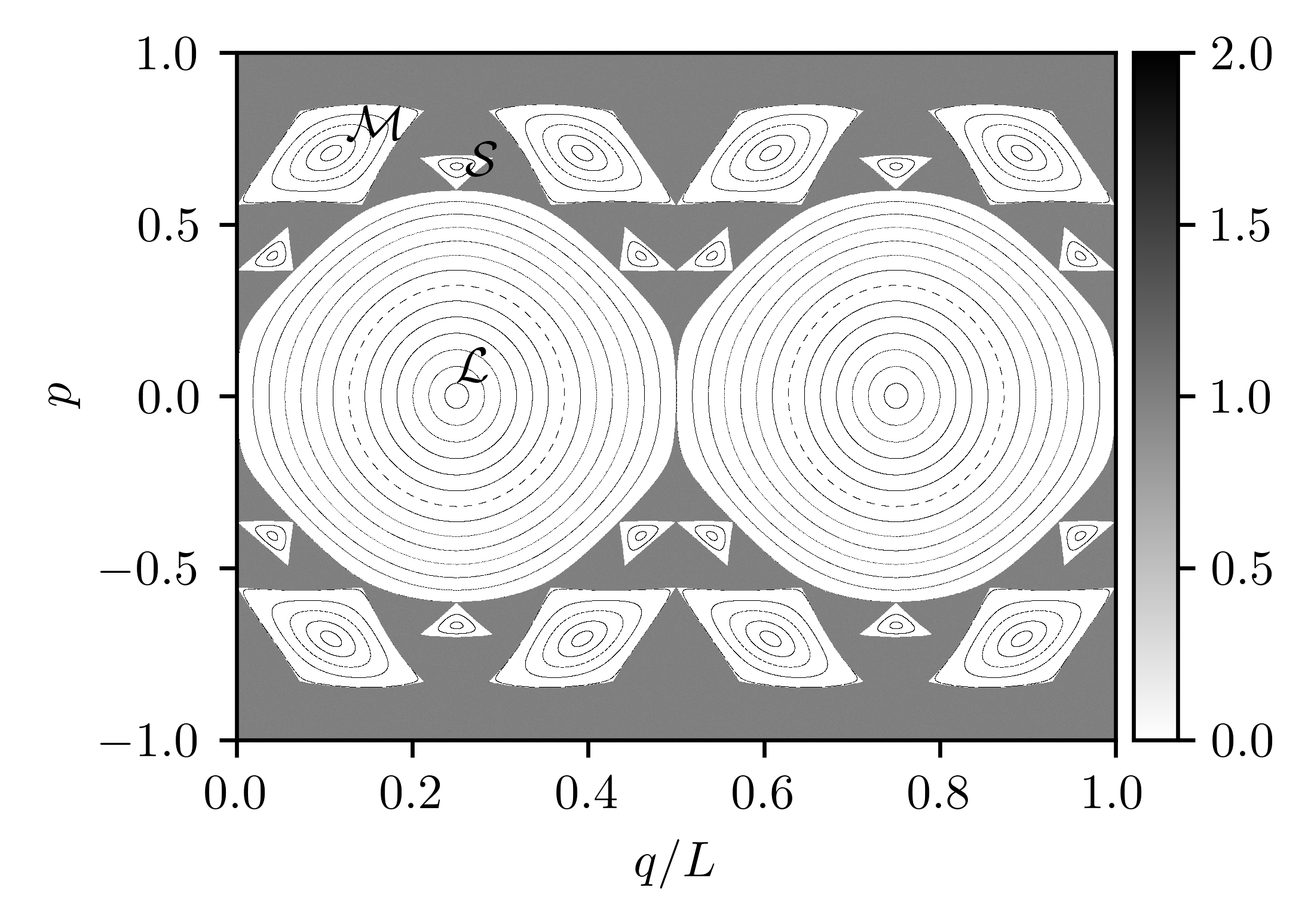}  
   \par\end{centering}
 \caption{ The phase portrait of the lemon billiard $B=0.1953$.
   The chaotic component is mapped by a single chaotic orbit and recurrence time statistics are presented as grayscale plot. The
   gray coding on the color bar is the quantity $S$ measuring the stickiness in the chaotic component (see Appendix \ref{AppendixB}), showing that we have uniformly $S\approx 1$, thus no significant stickiness. Inside the islands of stability we show some representative invariant tori. $\chi_c=0.3585$, and $\rho_c=0.2804$, $\rho_r=1-\rho_c=0.7196$ and ${L}=5.4969$.}
\label{fig: s_plot_1}
\end{figure}

\begin{figure}
 \begin{centering}
   \includegraphics[width=9cm]{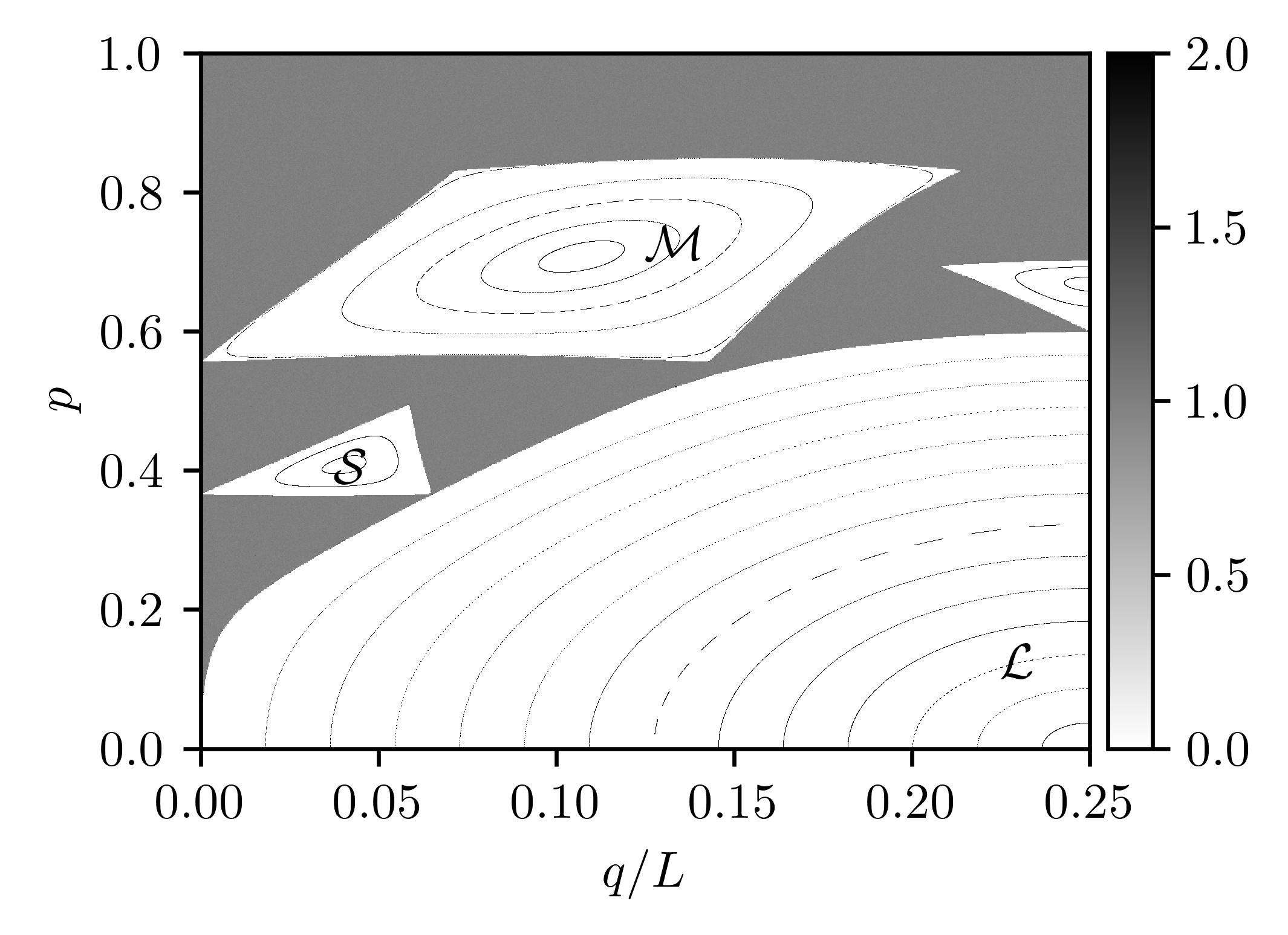}  
   \par\end{centering}
 \caption{ The details of the desymmetrized part of the phase portrait of the lemon billiard $B=0.1953$. For description, see Fig. \ref{fig: s_plot_1}.}
\label{fig: s_plot_2}
\end{figure}

The phase portrait for the billiard $B=0.1953$ is shown in Figs. \ref{fig: s_plot_1} - \ref{fig: s_plot_2}.
The relative fractions of chaotic component are $\chi_c=0.3585$ or $\rho_c=0.2804$ (which is the Berry-Robnik parameter).
Three independent regular island chains are clearly visible,
the largest one around the period-2 orbit which is
densely covered by the invariant tori, 
with no visible thin chaotic layers inside.  We denote
the largest island chain by $\cal L$, the second largest one by $\cal M$ and
the smallest one by $\cal S$. The relative phase space
volume of all three regular regions taken together is
$\rho_r=1-\rho_c= 0.7196$. The chaotic sea is
very uniform, with no significant stickiness regions,
as evident in the $S$-plot (introduced in Ref. \cite{Lozej2020}). The $S$-parameter is the local coefficient of variation of the recurrence times into small areas of the chaotic component. When $S=1$ the distribution of recurrence times is exponential, as expected for uniform chaos. If $S>1$ this indicates a modified recurrence time distribution, i.e. stickiness. For more details, see Appendix \ref{AppendixB}.

\begin{figure}
 \begin{centering}
    \includegraphics[width=9cm]{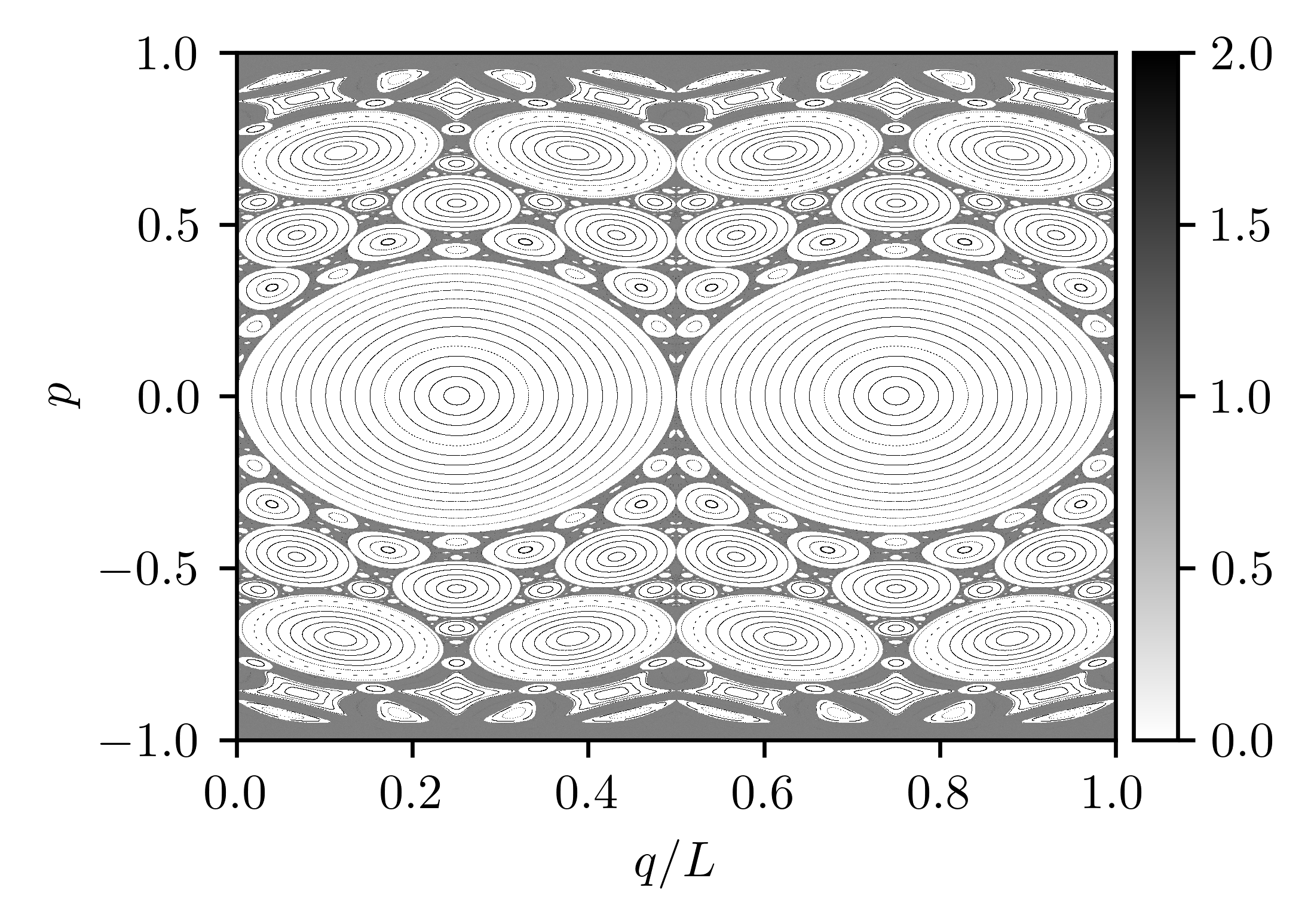} 
   \par\end{centering}
 \caption{ The phase portrait of the lemon billiard $B=0.083$. For description, see Fig. \ref{fig: s_plot_1}.
   $\chi_c=0.2168$, and $\rho_c=0.1617$,
   $\rho_r=1-\rho_c=0.8383$, ${L}=5.9508$.
    }
\label{fig: s_plot_desym_1}
\end{figure}

\begin{figure}
 \begin{centering}
    \includegraphics[width=9cm]{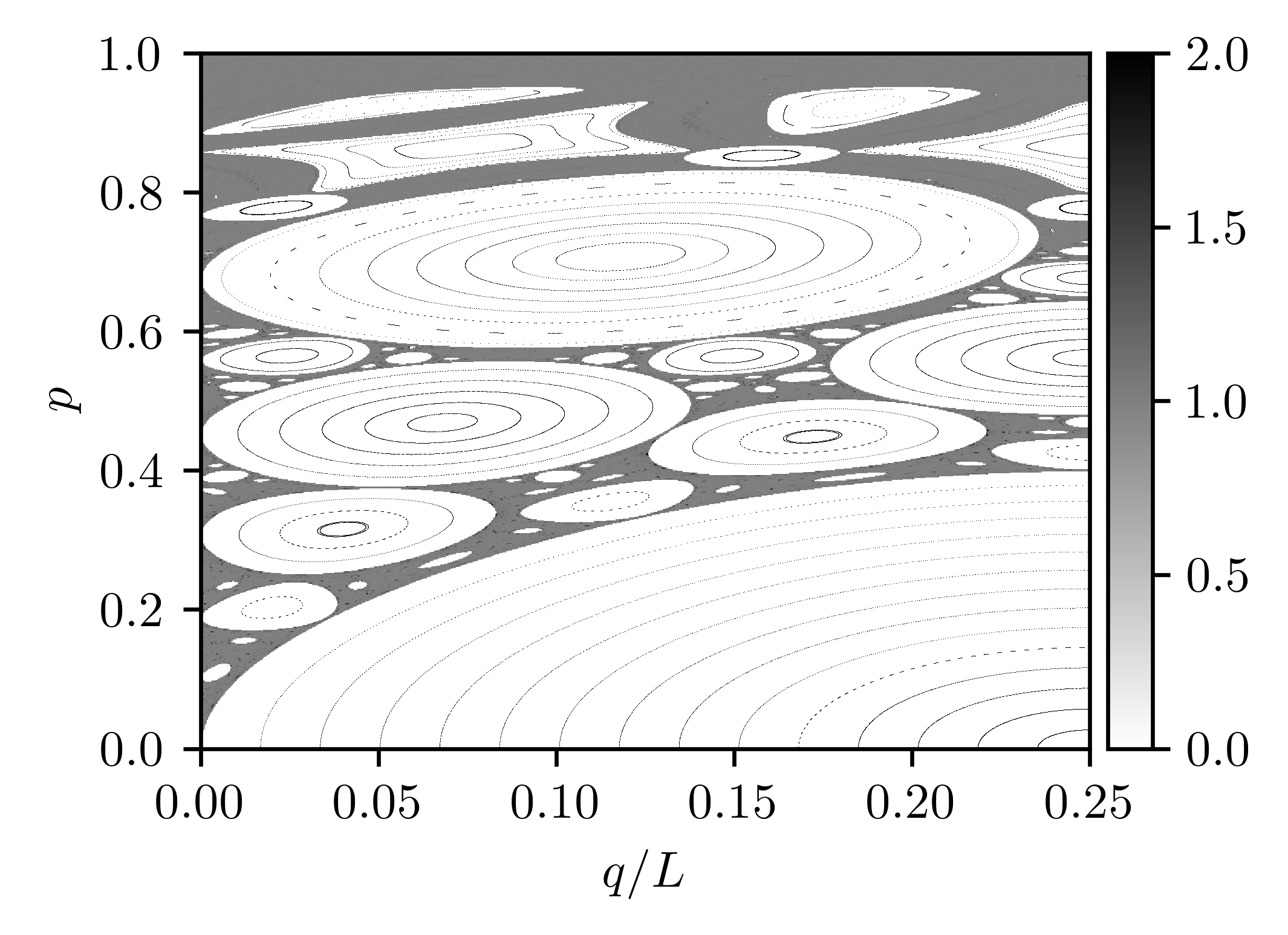} 
   \par\end{centering}
 \caption{The details of the desymmetrized part of the phase portrait of the lemon billiard $B=0.083$. For description, see Fig. \ref{fig: s_plot_1}.}
\label{fig: s_plot_desym_2}
\end{figure}

The phase portrait as shown in Fig. \ref{fig: s_plot_desym_1}
for the billiard $B=0.083$ is more complex.
The relative fraction of the area of the chaotic component
of the bounce map is $\chi_c=0.2168$, while the relative
fraction of the  phase space volume of the same chaotic
component is $\rho_c=0.1617$. Thus, the relative phase space
volume fraction of the complementary regular
regions is $\rho_r=1-\rho_c=0.8383$. Also in this case the
chaotic sea is rather uniform, with no significant stickiness regions.
The details of the desymmetrized part of the  phase space are shown in Fig. \ref{fig: s_plot_desym_2}.

We can conclude that the two cases $B=0.1953,\;0.083$ are
interesting to verify the Berry-Robnik picture of quantum billiards,
including the possible quantum localization of the chaotic
eigenstates and the universal statistical properties of the localization measures. The two cases add a level of increased complexity for the island structure, while keeping the chaotic sea uniform, without stickiness as in the examples of the previous paper \cite{LLR2021A} where simple single-island phase spaces were studied. 
This is quite different from the ergodic case $B=0.5$ studied
in Ref. \cite{LLR2021}, where the stickiness effects are strongly
pronounced and are reflected in the nonuniversal localization properties of
quantum (chaotic) eigenstates.

\section{The Schr\"odinger equation and the
  Poincar\'e - Husimi functions}
\label{sec3}

The quantum billiard ${\cal B}$ is described
by the stationary Schr\"odinger equation, in the chosen units
($\hbar^2/2m=1$) given by the Helmholtz equation

\be \label{Helmholtz}
\Delta \psi + k^2 \psi =0
\ee
with the Dirichlet boundary conditions  $\psi|_{\partial {\cal B}}=0$.
The energy is $E=k^2$. 

The {\em mean number of energy levels}
${\cal N} (E)$ below $E=k^2$ is determined quite
accurately, especially at large energies, asymptotically exactly,
by the celebrated Weyl formula (with perimeter corrections)
using the Dirichlet boundary conditions, namely

\be \label{WeylN}
{\cal N} (E) = \frac{{A}\;E}{4\pi} - \frac{{L}\;\sqrt{E}}{4 \pi} + c,
\ee
where $c$ are small constants determined by the corners and the
curvature of the billiard boundary. Differentially, they play no role.
Thus, the density of states
$d (E) = d{\cal N}/dE$ is equal to

\be \label{Weylrho}
d(E) = \frac{{A}}{4\pi} - \frac{{L}}{8 \pi \sqrt{E}}.
\ee
As shown in the previous paper \cite{LLR2021B} and according to the
theoretical predictions by Steiner \cite{Steiner1994,SteinerPRL1994},
the fluctuations of the number of energy levels (mode-fluctuations)
around the mean value of Eq. (\ref{WeylN}) grow with $k$, such that their variance
increases linearly with $k$ in integrable systems and as $\log k$
in ergodic chaotic systems. For the mixed-type systems
it is something in between, namely the variance is the sum of the
variances of the regular and of the chaotic part, provided that
we can treat them as statistically independent of each other, which
in the semiclassical regime is a valid assumption according to the
Berry-Robnik picture \cite{BerRob1984}. 
Thus, the fluctuations at large values of $k$ in all cases
can be very large, and the standard deviation even diverges as
$k\rightarrow \infty$. In ergodic chaotic systems
the distribution of fluctuations is predicted to be Gaussian
\cite{Steiner1994,SteinerPRL1994}, while in Ref. \cite{LLR2021B}
we have shown that it is almost always Gaussian, or very close to that,
even in the integrable and mixed-type systems.

Our numerical method to compute the eigenfunctions
is based on the Vergini-Saraceno scaling method
\cite{VerSar1995,LozejThesis}, with two possible basis sets, plane waves or
circular waves (Bessel functions for the radial part and trigonometric
functions for the angular part). The numerical methods are available as part of a Python numerical library \cite{LozejThesis, QuantBill}. The agreement between calculations using both basis sets is good, so all the presented calculations were done using the plane wave basis because of the much faster computation.
The method computes several eigenstates within a small energy interval per diagonalization, and its efficiency allowed us to compute almost $10^6$ eigenstates in the PH representation for each billiard. However, the precision of the computed energy levels decreases with the distance from the center of the energy interval. Thus, even after careful comparison of the levels in overlapping energy intervals, errors in the accumulation of levels still occur, and some levels are lost. The number of missing levels was never larger than 1 per 1000 levels and due to the overall large number of eigenstates this should have very little effect on the statistical results.

The lemon billiards have two reflection symmetries, thus four symmetry classes:
even-even, even-odd, odd-even and odd-odd.  For the purpose of our analysis we have considered only the quarter billiard, i.e. the odd-odd symmetry class. We have calculated the energy spectra and PH functions for each billiard in spectral stretches of about $10^6$ states starting from the $10^4$-th state. In order to compare the energy dependent results between the two billiards we must first exclude the effect of densities of states. We use the standard unfolding procedure and insert the spectra into the Weyl formula (\ref{WeylN}) 
\begin{equation}
    e_n := {\cal N}(E_n),
\end{equation}
where $e_n$ is the unfolded energy of the n-th state. This results in a spectrum where the mean level spacing is equal to one.

As in previous works \cite{BatRob2013B,BLR2018, BLR2019B,BLR2020,LLR2021,LLR2021A,LLR2021B}
we now define the Poincar\'e - Husimi functions (PH functions),
thereby introducing the quantum phase space whose structure
should correspond to the classical phase space in the
semiclassical limit.
Thus, instead of studying the eigenstates by means of the wavefunctions
$\psi_m({\bf r})$ as solutions of the Helmholtz
equation (\ref{Helmholtz}) we define  PH
functions as a special case of Husimi functions \cite{Hus1940},
which are in turn Gaussian smoothed Wigner functions \cite{Wig1932}.
They are very natural for billiards. Following Tuale and Voros
\cite{TV1995} and B\"acker et al \cite{Baecker2004} we define the properly
${L}$-periodized coherent states
centered at $(q,p)$, as follows

\bea \label{coherent}
c_{(q,p),k} (s) & =  & \sum_{m\in {\bf Z}} 
\exp \{ i\,k\,p\,(s-q+m\;{L})\}  \times \\ \nonumber
 & \exp & \left(-\frac{k}{2}(s-q+m\;{L})^2\right). 
\eea

The Poincar\'e-Husimi function is then defined as the absolute square
of the projection of the boundary function $u_m(s)$ onto the coherent
state, namely

\be \label{Husfun}
H_m(q,p) = \left| \oint c_{(q,p),k_m} (s)\;
u_m(s)\; ds \right|^2.
\ee
where $u_m(s)$ is the boundary function, that is the normal derivative
of the eigenfunction of the $m$-th state
$\psi_m({\bf r})$ on the boundary at point $s$,

\be  \label{BF}
u_m(s) = {\bf n}\cdot \nabla_{{\bf r}} \psi_m \left({\bf r}(s)\right).
\ee
Here ${\bf n}$ is a unit outward normal vector to the
boundary at point ${\bf r}(s)$.
The boundary function satisfies an integral equation and
also uniquely determines the value of the wavefunction $\psi_m ({\bf r})$
at any interior point ${\bf r}$ inside the billiard ${\cal B}$.

According to the principle of uniform semiclassical condensation (PUSC) of
the Wigner functions and Husimi functions (see \cite{Rob1998,Rob2020}
and the references therein) 
the PH functions are expected to condense (collapse) in the semiclassical
limit either on an invariant torus or on the chaotic component in the
classical phase space. This is exactly what we observe -
the higher the energies the sharper the condensation/collapse
of the PH functions. At not sufficiently high energies {\em mixed-type
eigenstates (PH functions)} may exist due to the tunneling between
regular and chaotic domains (see \cite{Vidmar2007,BatRob2010}
and references therein), but their number is expected to
decrease monotonically with increasing $e$.

\section{Separating regular and chaotic eigenstates}
\label{sec4}

We use the PH functions to identify regular and chaotic eigenstates,
simply by the criterion of overlap with the classical invariant tori
or the chaotic region, respectively. This has been introduced and
implemented in our previous papers \cite{BatRob2013A,BatRob2013B},
see also Refs. \cite{Rob2016,Rob2020}. There we have defined an
overlap index $M$, which in ideal case is $+1$ for chaotic states
and $-1$ for regular states. Namely, we discretize the classical
phase space $(q,p)$ and the quantum phase space defined by the
PH functions $H(q,p)$ into a rectangular grid of points indexed by
$(i,j)$ centered in cells of equal area,
and normalize the PH functions in such a way that
$\sum_{i,j}H_{i,j}=1$. At each grid point, we define a discrete
quantity $C_{i,j}$ such that it is $+1$ if the grid point $(i,j)$
belongs to the chaotic region, and $-1$ otherwise. The chaotic region
is constructed/generated by a single sufficiently long and
dense chaotic orbit. This implies
that the complement contains all the regular regions and possibly the
other smaller chaotic regions. Typically, these smaller chaotic
regions are so small that they can be neglected and treated as if they
belonged to the regular part. 

We calculate the overlap index $M$ as follows

\be \label{Mindex}
M = \sum_{i,j} H_{i,j}\; C_{i,j}.
\ee
Ideally, in the sufficiently deep semiclassical limit, $M$ should obtain
either exact value $+1$ or $-1$, for the chaotic or regular type of the
PH function, respectively. In practice, since the semiclassical limit is not yet achieved, $M$ assumes also values between $+1$ and $-1$.
The question arises, what value $M=M_s$ should be taken as the criterion to separate
the regular and chaotic eigenstates.
In the past \cite{BatRob2013A,BatRob2013B}
we have used two possible physical criteria, (i) the classical one,
and (ii) the quantum one. In the former case, we choose $M_s$
such that the fractions of regular and chaotic states are
the classical values $\rho_r$
and $\rho_c=1-\rho_r$, respectively. The quantum criterion for $M_s$  is
such that the fit of the chaotic level spacing distribution best
agrees with the Brody distribution. This method is applicable in
a general case. 
In Sec. \ref{sec6}  we will analyze the distributions of the overlap index in relation to the localization measures and show that the number of mixed-type
eigenstates (their PH functions) with intermediate values of $M$, monotonically decreases with increasing energy as expected in the Berry-Robnik picture.

\section{Localization measures}
\label{sec5}

Following our previous papers, e.g. \cite{BatRob2013A} (see also
\cite{LLR2021A}), we will now introduce the localization measures for PH functions. They are special cases of the more general localization measures based on Wehrl entropy \cite{wehrl1978entropy}, also recently studied in references \cite{villasenor2021, pilatowsky2022effective, PilatowskyCameo2022}. We define the {\em entropy localization measure of a single eigenstate - of its PH function - $H_m(q,p)$} as

\be \label{locA}
A_m = \frac{\exp I_m}{N_c},
\ee
where

\be  \label{entropy}
I_m = - \int dq\, dp \,H_m(q,p) \ln \left((2\pi\hbar)^f H_m(q,p)\right)
\ee
is the information entropy of the eigenstate labelled by $m$.
Here $f$ is the number of degrees
of freedom (for 2D billiards $f=2$, and for surface of section it is
$f=1$) and $N_c$ is a number of Planck's cells on the 
classical domain, $N_c=\Omega/(2\pi\hbar)^f$, where
$\Omega$ is the classical phase space volume. In the case of the
uniform distribution (extended eigenstates) $H=1/\Omega={\rm const.}$
the localization measure is $A=1$, while in the case
of the strongest localization $I=0$, and $A=1/N_C \approx 0$.
The Poincar\'e-Husimi function $H(q,p)$
(\ref{Husfun}) (normalized as mentioned before)
was calculated on the grid points $(i,j)$
in the phase space $(q,p)$,  and
we express the localization measure in terms of the discretized function
$H_{i,j}$ as follows

\be \label{AfromH}
A_m= \frac{1}{N} \exp\left( -\sum_{i,j}H_{i,j}\; \ln H_{i,j} \right),
\ee
where $N$ is the number of grid points of the rectangular
mesh with cells of equal area.
We have $H_{i,j}=1/N$ in the case of complete (constant)
extendedness, and $A=1$.  In case of maximal localization
we have $H_{i,j}=1$ at just one point, and zero elsewhere,
yielding $A=1/N \approx 0$ for large $N$. In all calculations we
have used the grid of $1000\times 750$ points, thus $N = 750000$.

According to this definition, regular eigenstates that condense on invariant tori are localized. The distribution of the localization measures $P(A)$ for these regular states is close to linear, starting from small $A\approx 0$ increasing up to a maximal cut-off value $A=A_c$ , corresponding to the outermost torus (last torus) of the regular island as explained in \cite{LLR2021A}. This applies to each possible chain of regular islands.

The chaotic PH functions can be either strongly localized or extended, but
never entirely uniformly extended (i.e. not uniformly constant),
as they experience oscillations and display
a characteristic pattern of their nodal (zero level) points
\cite{TV1995,Prosen1996}. Therefore, the maximal value of $A$, denoted by $A_0$, is approximately $A_0\approx 0.7$ according to empirical studies with real energy spectra \cite{LLR2021A}.
The random wavefunction model yields numerically
$A_0\approx 0.694$ \cite{LozejThesis}, while a theoretical estimate of
the random wavefunction model in the {\em ultimate semiclassical limit}
yields (see Refs. \cite{KusMosHaake1988, Jones1990, GunZyc2001, pilatowsky2021, PilatowskyCameo2022}) $A_0= e^{(\gamma -1)} \approx 0.65522$,
where $\gamma=2.71828$ is the Euler constant. In mixed type systems the upper bound of the measure will be further reduced compared to
fully chaotic ergodic systems, since not all the phase space is accessible.  One must renormalize/divide the measured
$A$ by the relative area $\chi_c$ of the chaotic component in the
phase space to make a quantitative comparison for the upper bound in the two different settings, as was done in our previous work. In this work we examine regular, chaotic and mixed states and will not rescale the measures of the chaotic states and keep the definition Eq. (\ref{locA})  for all types of states.

Of course, one may define many localization measures that are more sensitive to different features of the underlying PF functions.
In the literature, many possibilities are presented,
and the most general definition is based on the Renyi entropy
(see Refs. \cite{villasenor2021, PilatowskyCameo2022}) of class $\alpha$. $\alpha=1$ corresponds
to the information (Shannon) entropy on which the localization measure
$A$ in Eq. (\ref{locA}) is based. Another localization measure is
the normalized inverse participation ratio $R$ corresponding to
$\alpha=2$. Here we define it as follows in terms of the discretized
and normalized Poincar\'e-Husimi function:

\be  \label{nIPR}
R =  \frac{1}{ N \sum_{i,j} H_{i,j}^2}.
\ee
$R=1/N$, where $N$ - the number of grid points - is very large,
corresponds to the maximal localization
$R\approx 0$, while $R=1$ corresponds to the full extendedness
(delocalization, where $H_{i,j}=1/N$ for all $(i,j)$ ). At this point, it is useful to discuss differences between the measures $A$ and $R$ and develop some intuitive understanding. As already discussed, both measures are sensitive to the overall "size" or "extendedness" of the PH function in the phase space. However, taking a logarithm of the PH function as in $A$ will suppress the differences in the magnitudes between the high intensity and low intensity areas, whereas taking a square of the PH function further enhances the peaks where $H_{i,j}>1$ and suppresses the low intensity background where $H_{i,j}<1$. The measures will thus produce the most distinct results for PH functions that are strongly peaked in just a small area, but have a much larger support. This makes the measure especially suitable for finding states scarred by periodic orbits, as demonstrated for the Dicke model by Pilatowsky-Cameo et. al. in Ref. \cite{pilatowsky2021} and in triangular billiards \cite{lozej2022triangles}. 
In previous works \cite{BLR2019B,BLR2020} we have shown that after a local averaging in energy, the relationship between $A$ and $R$ for chaotic states is linear. Here we explore the relationship further by examining the joint probability distribution density $P(A,R)$, that is the probability of finding an eigenstate within an infinitesimal box $[A, A+dA] \times [R, R+dR]$. We normalize the distributions on the rectangle $(A,R)\in[0,0.4]\times[0,0.3]$.

\begin{figure*}

 \begin{centering}
    \includegraphics[width=16cm]{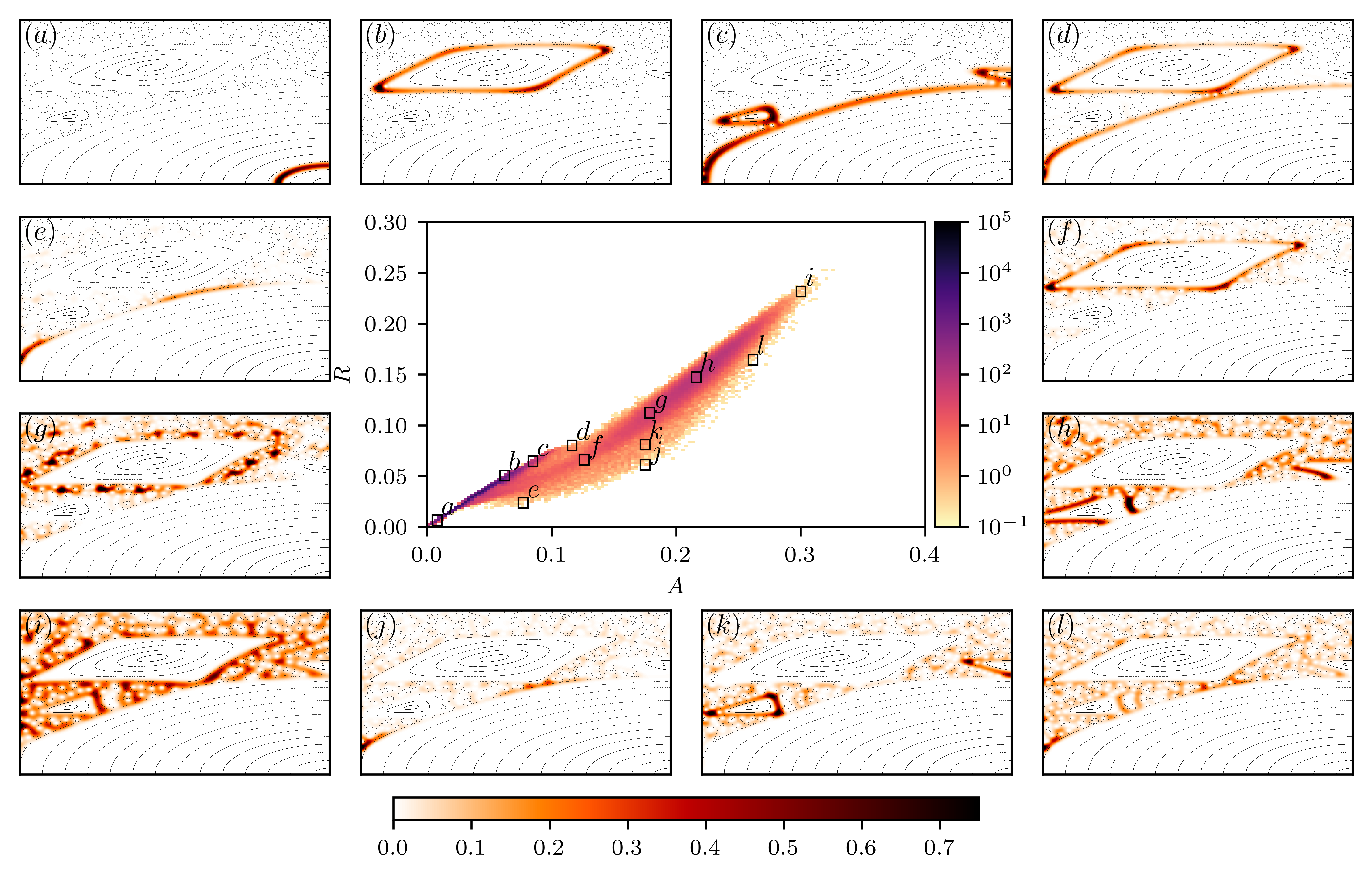}
   \par\end{centering}
 \caption{(central) color-plot of the histogram of the joint probability density $P(A,R)$ for approximately $10^6$ eigenstates with unfolded energy $e\in[10^4,10^6]$ of the $B=0.1953$ lemon billiard. The color scale of the main figure is logarithmic. PH functions of the highest energy eigenstates within small boxes at various positions are shown on the margin. Their corresponding wavenumbers are (a) $k=4638.3906$, 
(b) $k=4131.8411$, 
(c) $k=2839.9081$, 
(d) $k=3889.6615$, 
(e) $k=4609.8260$, 
(f) $k=4637.4242$, 
(g) $k=4636.6707$, 
(h) $k=4638.1850$, 
(i) $k=2813.2003$, 
(j) $k=4471.7931$, 
(k) $k=4629.2434$, 
(l) $k=4501.5252$. A classical phase portrait is plotted in the background of each state for comparison. The color scale at the bottom encodes the relative amplitude of the PH function.  }
\label{fig: measures_compare_1}
\end{figure*}

Let us first consider the $B=0.1953$ lemon billiard, with the comparatively simpler phase space featuring three stable island chains. In Fig. \ref{fig: measures_compare_1} we show the joint probability distribution density $P(A,R)$ as a colorplot together with some representative PH functions from different regions of the parameter space. The PH functions are selected as the highest energy eigenstate found in a local area of the plot. Let us first focus only on the distribution presented in the central figure. The white areas are regions where no eigenstates are found. We immediately notice the distribution is supported only on a small area near the diagonal of the rectangle. This corroborates the previously known result that the localization measures $A$ and $R$ are on average linearily related. We also observe two main clusters of eigenstates (note that the colorscale is logarithmic), a very sharp peak supported on a nearly one-dimensional line segment and a wider peak in the more delocalized regime. The first cluster is formed by the regular states localized on the invariant tori of the islands of stability, as can clearly be seen by examining some representative PH functions presented in (a) and (b). The second cluster is formed by the chaotic eigenstates, with representative examples (g), (h), (i), and (l) on the periphery. We will refer to the two clusters as the regular and chaotic cluster, respectively. They are visibly connected by additional structures. Outside the main clusters, many interesting mixed-states may be found. (c) and (d) show tunneling states between two of the outer tori of different island chains. These states are interesting, because they can support chaos assisted tunneling \cite{tomsovic1994tunneling}.The states (e), (f), (j), (k) represent chaotic states with a significant overlap with boundary tori, signifying tunneling between the chaotic and regular component. Although one would need to classify each state individually for a full description, the distribution together with the representative PH functions provide a good overview of the phenomenology of the eigenstates. One may also notice that some mixed states, e.g. (c) and (d), belong to much lower wavenumbers since states in this area cease to exist after a certain energy is reached. We will more thoroughly present the energy dependencies in Sec. \ref{sec7}.

\begin{figure*}
 \begin{centering}
    \includegraphics[width=16cm]{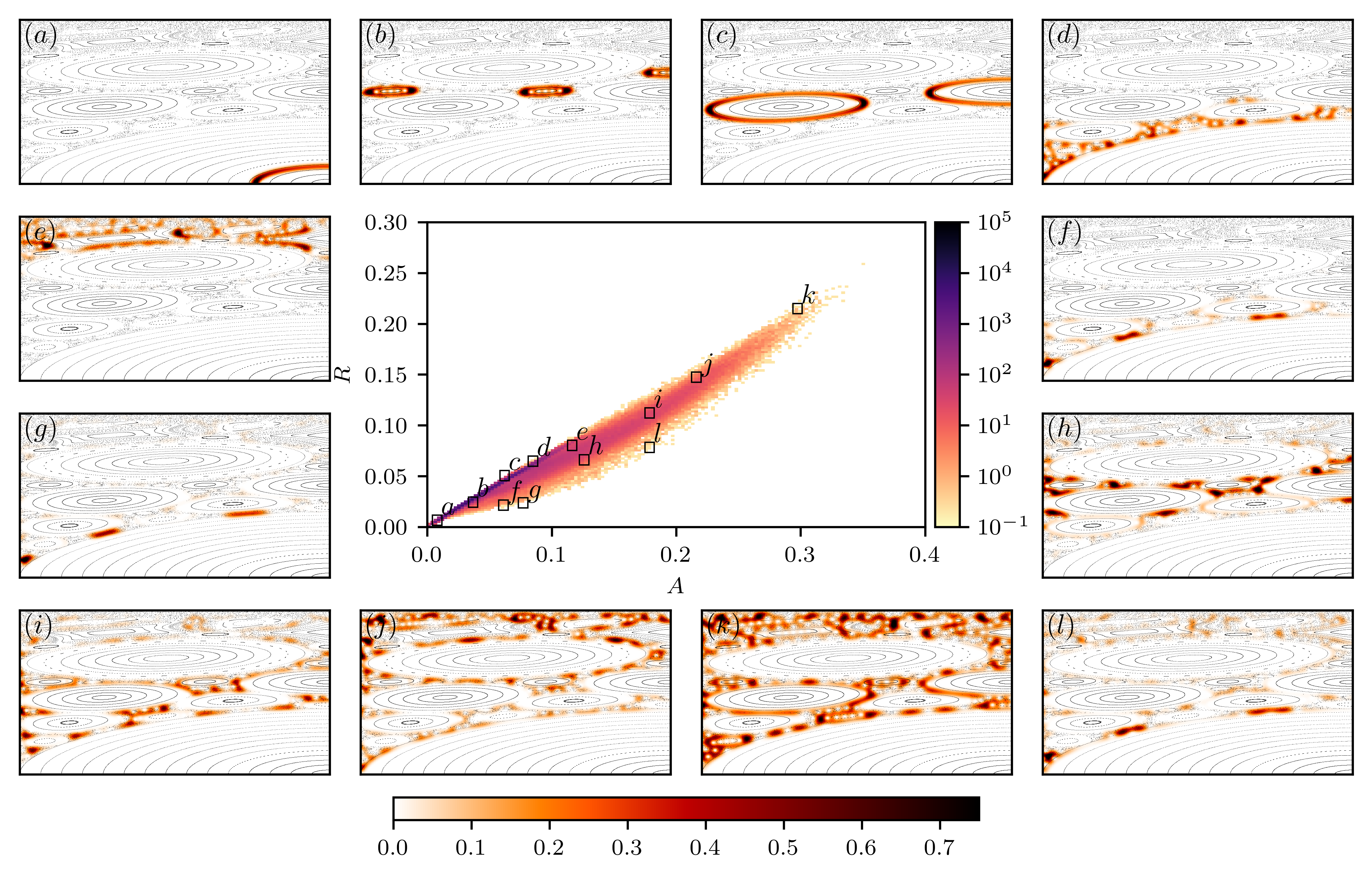}
   \par\end{centering}
 \caption{(central) color-plot of the histogram of the joint probability density $P(A,R)$ for approximately $10^6$ eigenstates with unfolded energy $e\in[10^4,10^6]$ of the $B=0.083$ lemon billiard. The color scale of the main figure is logarithmic. PH functions of the highest energy eigenstates within small boxes at various positions are shown on the margin. Their corresponding wavenumbers are (a) $k=4255.7099$, 
(b) $k=4255.6717$, 
(c) $k=4255.6550$, 
(d) $k=4253.5121$, 
(e) $k=4255.5289$, 
(f) $k=4251.1624$, 
(g) $k=4253.4432$, 
(h) $k=4255.0162$, 
(i) $k=4255.6096$, 
(j) $k=4254.0484$, 
(k) $k=3318.7134$, 
(l) $k=3927.8888$. A classical phase portrait is plotted in the background of each state for comparison. The color scale at the bottom encodes the relative amplitude of the PH function.}
\label{fig: measures_compare_2}
\end{figure*}

We now compare the results from the $B=0.083$ lemon billiard, with the by far more elaborate island structure, presented in the same manner in Fig. \ref{fig: measures_compare_2}. The same basic close to diagonal structure, with two main clusters, may be seen. However, the exact positions of the structures are different. This is not unexpected since the localization measures are sensitive to the geometry of the phase space and sizes of the tori, islands and chaotic component. The chaotic cluster is moved towards lower values of the localization measures, since the relative size of the chaotic component is smaller, but also because of dynamical localization effects that may be seen in the PH functions portrayed in (d), (e), (f). State (f) is also highly peaked near the boundary of the largest island of stability and might be interpreted as a mixed-state describing tunneling between the three nearby island chains. State (g) has strong peaks in the same areas, but extends weakly across the whole of the chaotic component. States (h), (i), (j), (k), (l) are extended chaotic states of various uniformity. However, some flooding into islands of stability is clearly visible. In particular, states (j) and (k) flood the topmost islands of stability completely and partially overlap other islands. Complete flooding of the topmost islands is also observed in the localized state (e). The states (a), (b) and (c) are regular eigenstates and are typical examples of states from the regular cluster.

To summarize, the joint probability distributions $P(A,R)$ are structurally similar in both example billiards. They feature two main clusters of states, the regular cluster supported on a very narrow strip (practically a line segment) and a wider chaotic cluster. The wide majority of states are found in these two clusters. The mixed states are found on the margins of the clusters and on additional system specific structure connecting the two main clusters that are related to the various tunneling processes between different island chains and the chaotic component. It is perhaps surprising that this much insight into the structure of the phase space may be gained by studying the relation between two simple localization measures of the same object, the PH function. The major advantage of this approach is that no prior information of the classical phase space is necessary. This may be of vital importance when studying higher dimensional systems, where the classical computations and producing detailed phase space portraits become increasingly difficult. In particular, the regular states are very easily identifiable since they cluster on such a narrow part of the parameter space.

\section{Overlap index and localization measures}
\label{sec6}

The classical computations presented is Sec. \ref{sec2} enable us to easily compute the overlap index of the classical chaotic and regular components and the PH functions, Eq. (\ref{Mindex}). This is information about the relative overlap with the classical structures, but this alone is not sufficient to identify the processes leading to only partial overlap. We therefore compare the overlap index of each individual state and its localization measures and study the joint probability distributions $P(A,M)$ and $P(A,R)$. We will focus on the distributions $P(A,M)$, that is, the probability of finding a state within an infinitesimal box $[A, A+dA] \times [M, M+dM]$. We normalize the distributions on the rectangle $(A,M)\in[0,0.4]\times[-1,1]$.

\begin{figure*}
 \begin{centering}
    \includegraphics[width=16cm]{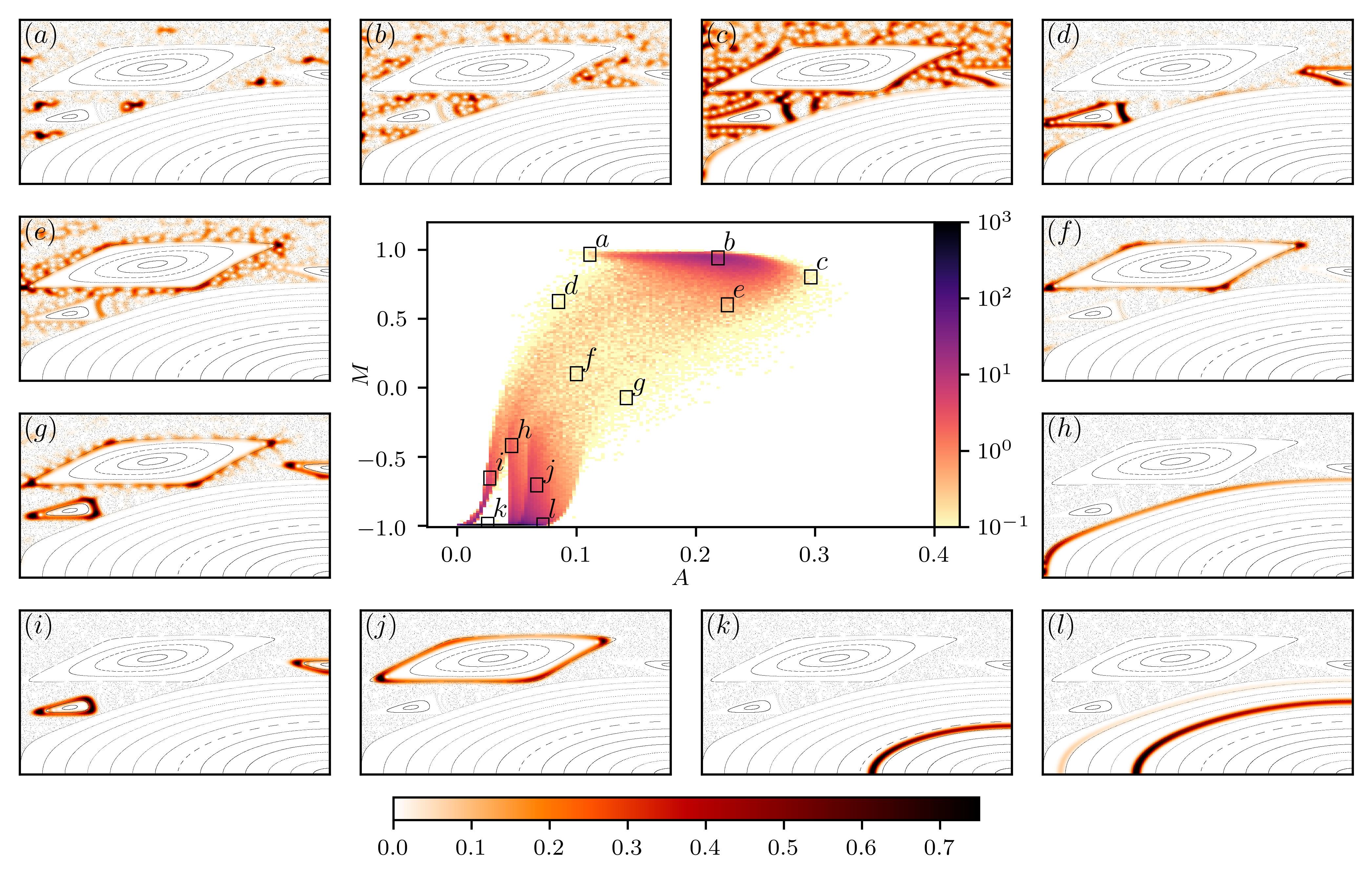}
   \par\end{centering}
 \caption{(central) color-plot of the histogram of the joint probability density $P(A,M)$ for approximately $10^6$ eigenstates with unfolded energy $e\in[10^4,10^6]$ of the $B=0.1953$ lemon billiard. The color scale of the main figure is logarithmic. PH functions of the highest energy eigenstates within small boxes at various positions are shown on the margin. Their corresponding wavenumbers are (a) $k=4598.5120$, 
(b) $k=4638.3417$, 
(c) $k=3486.3962$, 
(d) $k=4637.4426$, 
(e) $k=4610.3436$, 
(f) $k=4603.9386$, 
(g) $k=4050.5505$, 
(h) $k=4636.9990$, 
(i) $k=4637.1017$, 
(j) $k=4236.8109$, 
(k) $k=4638.4472$, 
(l) $k=3776.8166$. A classical phase portrait is plotted in the background of each state for comparison.The color scale at the bottom encodes the relative amplitude of the PH function.}
\label{fig: overlap_1}
\end{figure*}

In Fig. \ref{fig: overlap_1} we show the joint probability distribution density $P(A,M)$ in the $B=0.1953$ as a colorplot together with some representative PH functions from different regions of the parameter space. The PH functions are selected as the highest energy eigenstate found in a local area of the plot, and the color scale is logarithmic. The regular states belonging to the inner invariant tori form a sharp cluster at $M=-1$, extending to $A\approx 0.075$. An example is shown in (k) as well as an extreme example with tunneling between two invariant tori in (l). The chaotic states form a larger cluster with $M\gtrsim0.8$. Clearly, quite many chaotic states still have a small overlap with the outer tori of the regular islands, but can nevertheless be interpreted as purely chaotic states. On the $A$ axis, the chaotic cluster extends over a relatively large range from $A\approx0.1$ to $A\approx0.3$. The states (a), (b) and (c) show the transition from localized to increasingly uniform chaotic states. The regular and chaotic clusters are connected continuously by the mixed states. Going from the chaotic towards the regular cluster, we see various tunneling processes like for instance between the outer tori and the chaotic component (d), (e), (f) and also including outer tori of different island chains (g). In the lower part of the diagram $M <-0.5$, we see three structures of greater density. They correspond to states condensed on the boundary tori of the three island chains $\cal L$, $\cal S$ and$\cal M$ as is evident from the PH functions (h), (i) and (j), correspondingly. These states should be classified as regular. We have thus shown that the joint probability distribution $P(A,M)$ gives an excellent phenomenological overview of this mixed-type system.

\begin{figure*}
 \begin{centering}
    \includegraphics[width=16cm]{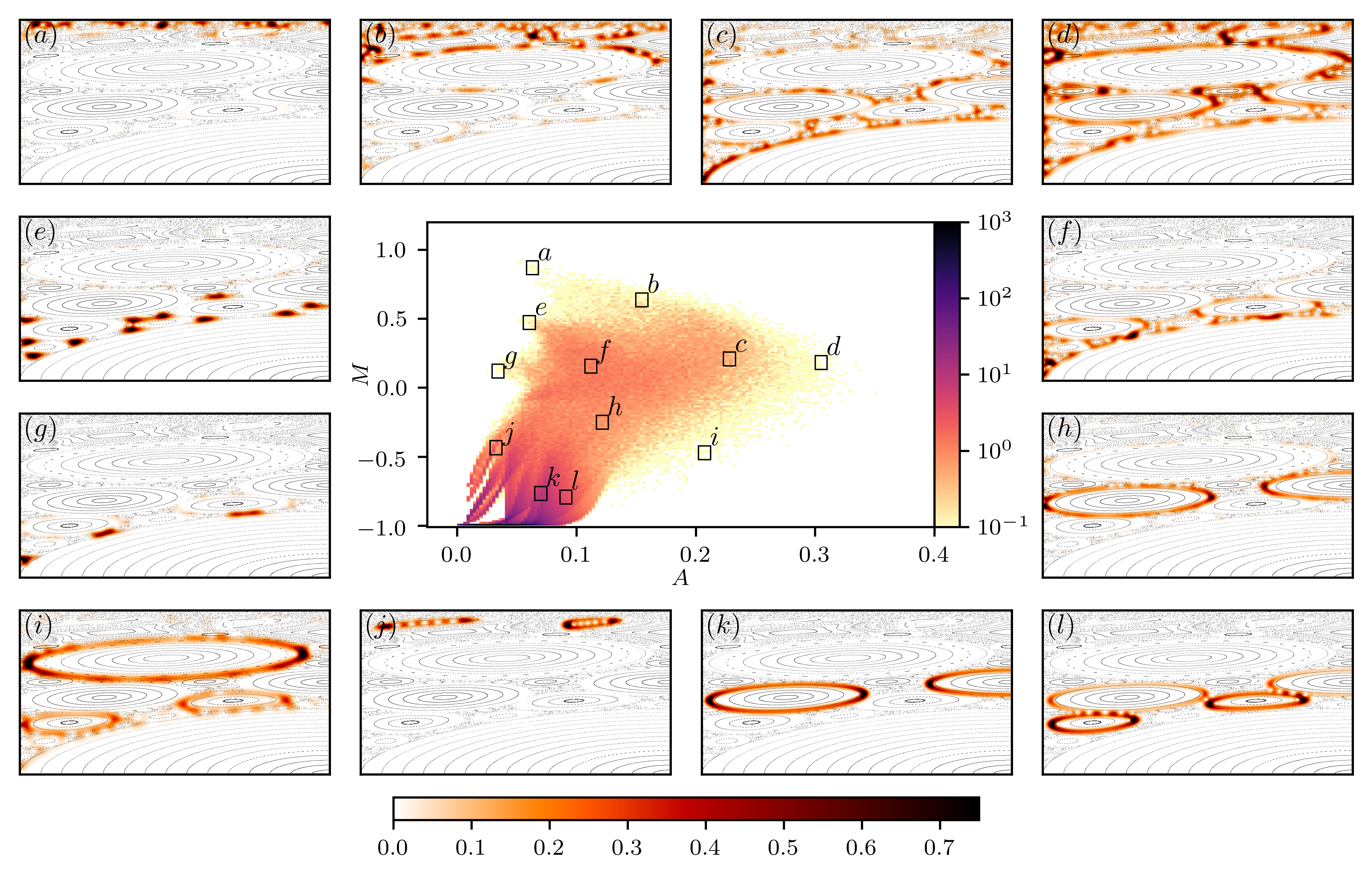}
   \par\end{centering}
 \caption{(central) color-plot of the histogram of the joint probability density $P(A,M)$ for approximately $10^6$ eigenstates with unfolded energy $e\in[10^4,10^6]$ of the $B=0.083$ lemon billiard. The color scale of the main figure is logarithmic. PH functions of the highest energy eigenstates within small boxes at various positions are shown on the margin. Their corresponding wavenumbers are (a) $k=4221.4610$, 
(b) $k=4253.9880$, 
(c) $k=4227.7699$, 
(d) $k=2649.8961$, 
(e) $k=4250.2980$, 
(f) $k=4254.5877$, 
(g) $k=4113.9895$, 
(h) $k=4254.0666$, 
(i) $k=2126.9352$, 
(j) $k=4255.2063$, 
(k) $k=4255.6550$, 
(l) $k=4181.7921$. A classical phase portrait is plotted in the background of each state for comparison. The color scale at the bottom encodes the relative amplitude of the PH function.}
\label{fig: overlap_2}
\end{figure*}

The Fig. \ref{fig: overlap_2} shows the distribution $P(A,M)$ in the other lemon billiard $B=0.083$. It is evident that the separation based on the values of $M$ is not so simple. The billiard $B=0.083$ has a much more complex classical phase portrait (see Figs. \ref{fig: s_plot_desym_1}, \ref{fig: s_plot_desym_2}). While the purely regular states on the inner invariant tori again form a sharp cluster at $M=-1$, the chaotic states are quite uniformly distributed over a wider range of $M$ values. Similarly to the previous example, the regular states of the outer tori of the different island chains form higher density "ridges" in the distribution. Because of the complexity of the phase space, there are now many of them, each corresponding to a chain of islands. Some of these outer tori states may be seen in examples (j) and (k) as well as (l) where we see tunneling between two nearby island chains. Progressing upwards towards the chaotic regime, states (h) and (i) exhibit tunneling between the outer tori and the chaotic sea. Crossing over into the predominantly chaotic regime $M>0$, we find some very localized states (a), (e), (f), (g) and some increasingly extended states (b), (c) and (d). The relatively small values of the overlap index are caused mainly by the flooding into the islands of stability, seen already in the joint localization measure distributions for this billiard ($B=0.083$). The mixed-type states are still strongly represented, their number
decreases with increasing energy $e$, but they would disappear only
at much higher energies. Purely chaotic states are practically non-existing in this
energy range, although their expected relative fraction is classically $\rho_2=0.1617$.

\begin{figure}
 \begin{centering}
    \includegraphics[width=8cm]{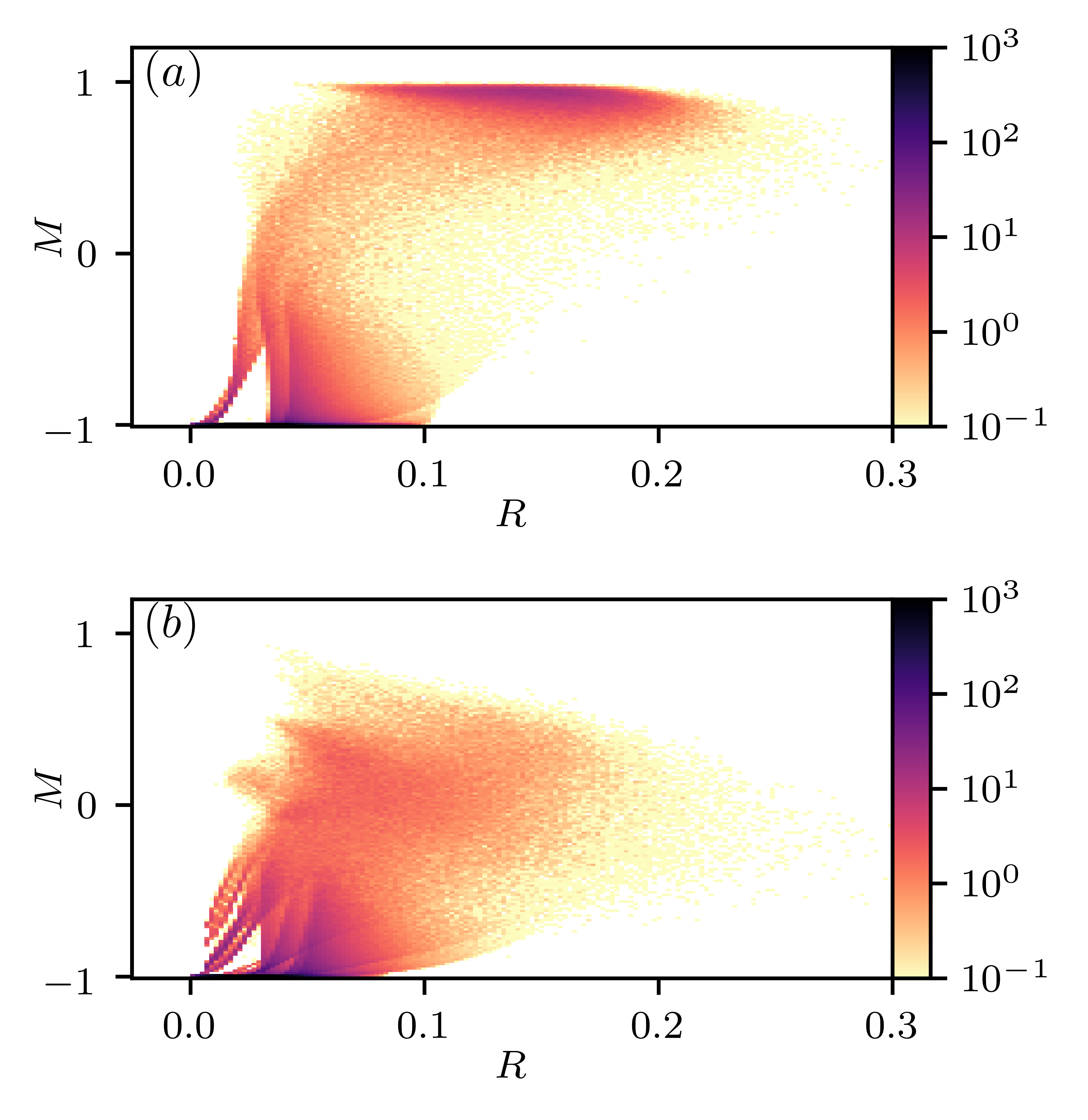}
   \par\end{centering}
 \caption{(central) color-plot of the histogram of the joint probability density $P(R,M)$ for approximately $10^6$ eigenstates with unfolded energy $e\in[10^4,10^6]$ of the (a) $B=0.1953$ and  (b) $B=0.083$ lemon billiard. The color scale is logarithmic.}
\label{fig: compare_overlap}
\end{figure}

Although there are subtle differences between the localization measures $A$ and $R$ (analyzed in the previous section) the general behavior of both is the same. For comparison, we show the distributions $P(R,M)$ in Fig. \ref{fig: compare_overlap}. 

\section{Energy dependence and the semiclassical condensation}
\label{sec7}

\begin{figure*}
 \begin{centering}
    \includegraphics[width=16cm]{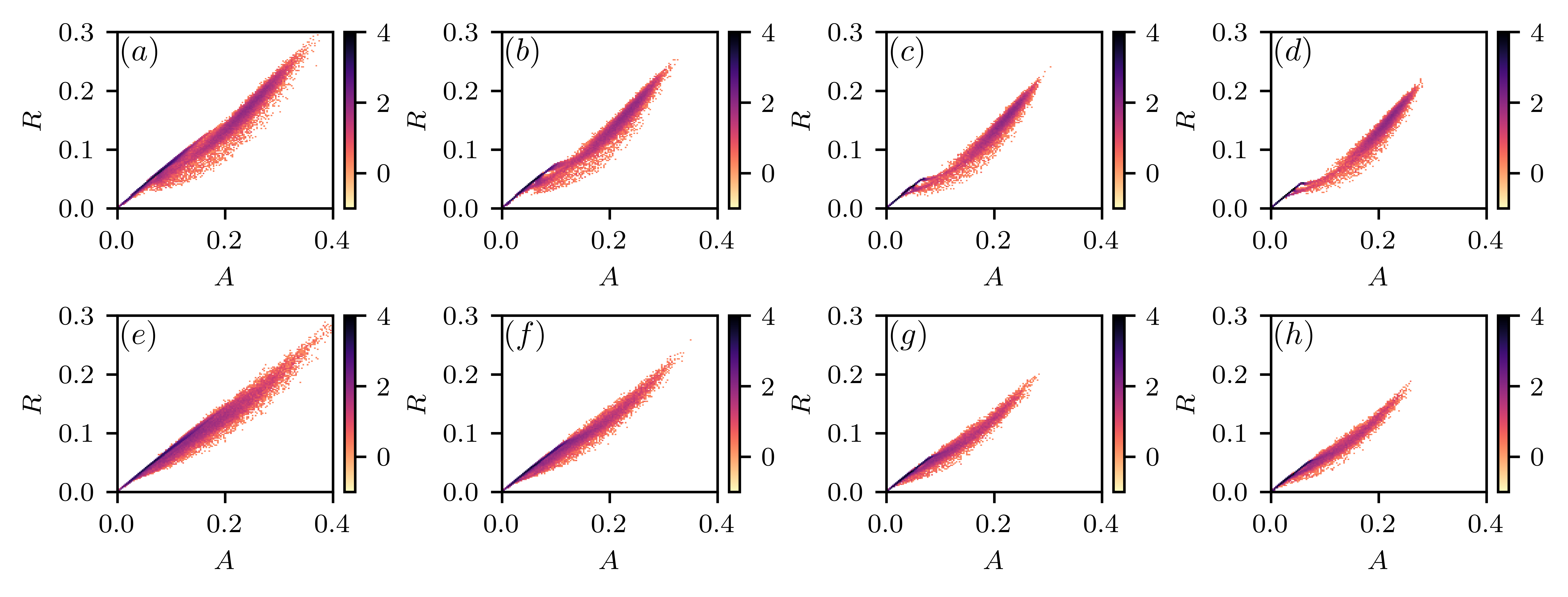}
   \par\end{centering}
 \caption{Color-plots of the histogram of the joint probability density $P(A,R)$ for approximately $10^5$ eigenstates for progressively higher energy intervals starting from (from left to right)  $e_0=10^4,10^5,5\times10^5,9\times10^5$ for the $B=0.1953$ (top row) and $B=0.083$ (bottom row) lemon billiard. The color scale is logarithmic. }
 \label{fig: measures_compare_energy}
\end{figure*}

In the semiclassical limit we expect the mixed chaotic regular states will gradually disappear in keeping with Berry-Robnik picture and PUSC. To study this, we consider the localization measures and overlap indices of eigenstates in narrower energy intervals, starting at progressively higher energies. Let us first inspect how the joint localization measure distributions $P(A,R)$ change with increasing the energy and progressing deeper into the semiclassical limit. This is shown in Fig. \ref{fig: measures_compare_energy} for both billiards. The invariant tori are one dimensional objects in the phase space.  When the energy is increased, the PH functions of the regular states condense on the invariant tori and become ever thinner and the localization measures of the regular states are decreased. The chaotic sea on the other hand is a positive measure set. In the final semiclassical regime, we would expect to see a delta distribution like peak at $(A,R)=(0,0)$ containing $\rho_1$ (the relative classical Liouville measure of the regular components) and another peak containing the $\rho_2=1-\rho_1$ chaotic states at $A=A_0$ and $R=R_0$. The exact position of the chaotic peak at ($A_0$, $R_0$) will depend on the measure of the chaotic component and the geometry of the phase space. In the $B=0.1953$ billiard (top row of Fig. \ref{fig: measures_compare_energy}), the chaotic and regular cluster are distinguishable already in the lowest energy interval, starting with $e=10^4$, although there are still many intermediate mixed states forming a smooth transition between them (we must keep in mind that the color scales are logarithmic and as such reduce the contrast between the different orders of magnitude). As we increase the starting energy, the mixed states gradually disappear, the features of the distribution $P(A,R)$ become sharper and a gap between the regular and chaotic states starts to open. The localization measures of the chaotic states (the center of the chaotic cluster), remain roughly the same with increased energy. Even though the semiclassical limit, with no mixed regular-chaotic states is not yet achieved, the trend towards this regime is evident. The joint distributions for the $B=0.083$ billiard are still further away from the semiclassical regime, owing to the greater complexity of the phase space. The center of the chaotic cluster is located at lower values of the localization measure, since the chaotic sea of this system is very thin. Flooding effects and dynamical localization are also present, as we observed in the representative PH functions in Sections \ref{sec5} and \ref{sec6}. All this contributes to the fact that only a slight indication of a gap opening between the chaotic and regular states is visible in the color plots.

\begin{figure*}
 \begin{centering}
    \includegraphics[width=16cm]{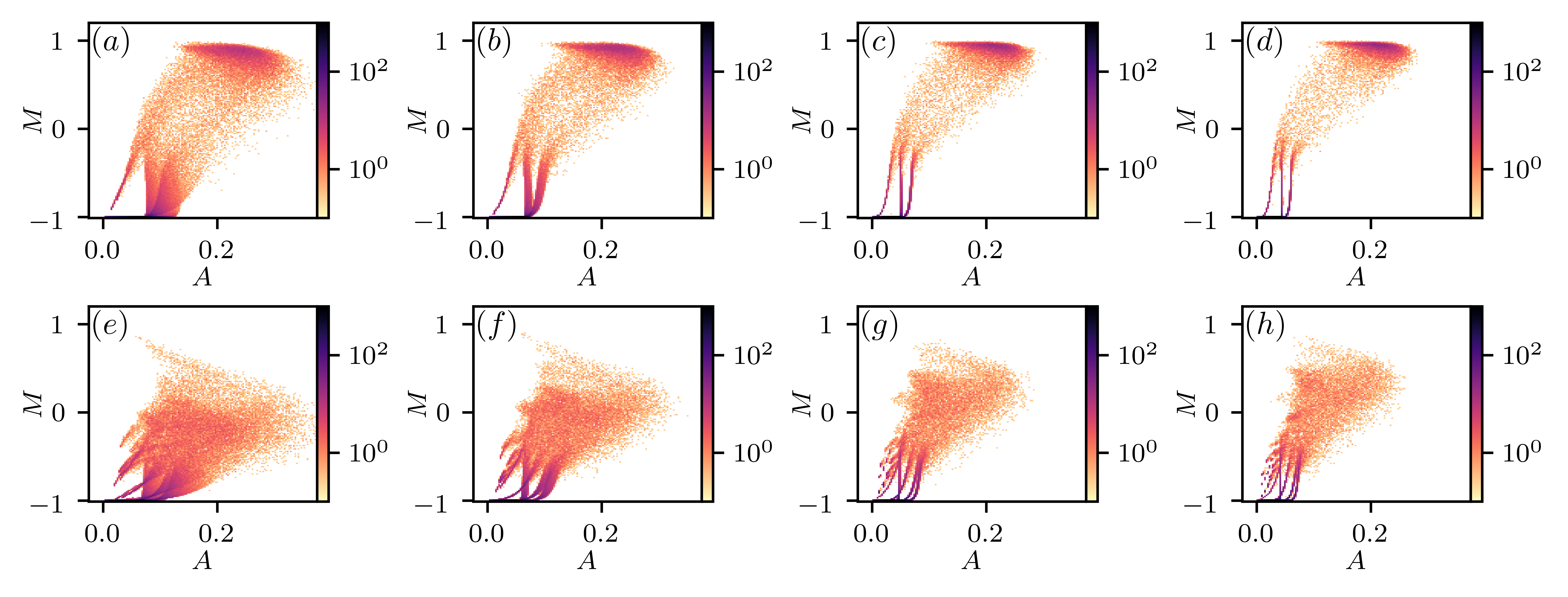}
   \par\end{centering}
 \caption{Color-plots of the histogram of the joint probability density $P(A,M)$ for approximately $10^5$ eigenstates for progressively higher energy intervals starting from (from left to right)  $e_0=10^4,10^5,5\times10^5,9\times10^5$ for the $B=0.1953$ (top row) and $B=0.083$ (bottom row) lemon billiard. The color scale is logarithmic. }
 \label{fig: measures_scatter_energy}
\end{figure*}

\begin{figure}
 \begin{centering}
    \includegraphics[width=9cm]{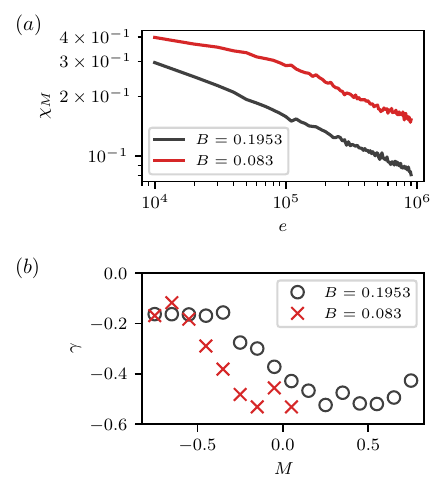}
   \par\end{centering}
 \caption{Decay of the relative number of mixed states with energy. (a) The relative number of states in the interval as a function of unfolded energy. Both decay exponents are close to $\gamma=-0.29$. (b) Decay exponents for smaller intervals $[M,M+\delta M]$ with $\delta M=0.1$.}
\label{fig: mixed_decay}
\end{figure}

The semiclassical condensation is even more evident when observing the energy dependence of the joint probability distributions $P(A,M)$. The distributions for different energy intervals are presented in Fig. \ref{fig: measures_scatter_energy}. In the $B=0.1953$ billiard, separation of the chaotic and regular eigenstates is very clear, and the mixed states become ever scarcer as we increase the energy. The three structures containing the outer tori of the three island chains also become very sharply defined. In accordance with our previous findings, the separation of the chaotic and regular states in the $B=0.083$ billiard is more ambiguous. However, we can still very clearly see the condensation of the states on the outer invariant tori of the many island chains and that mixed states, especially at $M<0$ appear less abundant. To quantify the decay of mixed states, we consider the following quantity. We define some interval in $M$ that we believe corresponds to the mixed states. Although this interval is somewhat arbitrary, the qualitative analysis of the PH functions may give an informed opinion on what values to take in each case. We then take an energy interval of width $w$ starting at $e$ and count the number of mixed states and divide by the number of all states in the interval. We observe the decay of this relative proportion of mixed states labeled $\chi_M(e)$  as a function of energy and also the interval taken in $M$. We find that for mixed states this quantity decays asymptotically as a power law $\chi_M(e)\propto e^{\gamma}$, where the exponent $\gamma<0$ depends on the interval we take in $M$. From the visual inspection of the PH functions (see Figs. \ref{fig: overlap_1} and \ref{fig: overlap_2}), we determined the mixed states are contained in the interval $M\in[-0.8,0.8]$ for the $B=0.1953$ and $M\in[-0.8,0.1]$ for the $B=0.083$ billiard. We found taking $w=10^4$ gives us a good compromise between the energy resolution and statistical significance for the relative number of states contained in the interval. In Fig. \ref{fig: mixed_decay} (a) we show the decay of the relative number of mixed states when taking the maximum interval in $M$ for each billiard. The decay rate is similar for both billiards $\gamma=-0.29$. Similar power law decays may also be seen when considering smaller intervals in $M$. In (b) we show the change of the decay exponent when taking relatively small intervals $[M,M+\delta M]$ with $\delta M=0.1$. The decay exponents show a transition from $\gamma \approx -0.15$ to $\gamma \approx -0.5$ with increasing $M$.  This indicates that the number of states related to the flooding processes decays faster than that of the states related to the tunneling from regular islands. The transition is similar in both billiards, but is slightly displaced in the $B=0.083$ case. 

\section{Summary, discussion and conclusion}
\label{sec8}

In the present paper, we have investigated two lemon billiards
with complex classical phase portraits of the mixed type, exhibiting
a dominant uniform classical chaotic component and several chains
of regular islands. The choice $B=0.1953$ and $B=0.083$ is based on the
systematic survey of the large family of lemon billiards in
Ref. \cite{Lozej2020}. The two billiards were singled out due to the complexity of the phase space and the absence of any apparent domains of stickiness as shown by the classical computations of recurrence time statistics. Our study is focused on the quantum mechanics of these billiards. In the present paper, we have presented a detailed analysis of the eigenstates in the Poincaré-Husimi representation and a phenomenological study, which gives a very good overview of all types of eigenstates, from regular to chaotic and mixed.

To summarize our main results: (i) We analyze the joint probability distributions of Reny-Wehrl localization measures of different orders of the eigenstates, revealing characteristic structures related to the regular chaotic and mixed eigenstates and quantifying their prevalence. (ii) We analyze the joint probability distributions of Reny-Wehrl localization measures in relation to a simple overlap measure with the classical phase space. The structure of the joint probability distribution allows for an even easier interpretation of chaotic mixed and regular states. Different tunneling processes may be identified, for instance: states on the boundary of the islands of stability (last invariant tori), states that support chaos assisted tunneling between islands, chaotic states that flood into islands of stability etc. (iii) We analyze the transition into the Berry-Robnik regime as a function of energy and show the fraction of mixed eigenstates decreases as a power law. The exponent depends on the type of tunneling process and ranges from -0.2 to -0.5.

A preliminary survey of the spectra of the two lemon billiards was published in Ref. \cite{LLR2021B}, where the fluctuation of the number
of the energy levels (mode fluctuation) was shown to obey quite well
the Gaussian distribution. The level spacing distribution of
the entire spectrum was shown to follow the
Berry-Robnik-Brody (BRB) distribution. In the billiard $B=0.1953$
the value of the level repulsion exponent $\beta$
(Brody parameter) is close to 1, reflecting the absence of dynamical
localization of the chaotic eigenstates, and the Berry-Robnik
parameter $\rho_r$ is close to its classical value. 
This is in line with the results of the current paper, where we see a very clear separation of the eigenstates at high energies in the same billiard.
On the other hand, in the billiard $B=0.083$, which has a much more
complex phase space structure, the results for $\beta$ and $\rho_r$
fluctuate significantly from case to case of four parities. A decrease of the level repulsion exponent has been linked to the presence of dynamically localized chaotic states \cite{BLR2018,BLR2019B} which we clearly observe in this billiard.
The fluctuations of the Berry-Robnik parameter can also be explained, since the energy dependent joint probability distributions of the localization and quantum-classical overlap measures show the asymptotic regime is not yet reached.

In the present paper, we presented a novel approach to interpret the localization measures by studying the joint probability distributions. 
We introduced two localization measures of individual PH functions, the entropy localization measure $A$ and the
normalized inverse participation ratio $R$, and the
overlap index $M$, which measures the degree of overlap
of the PH function with the regular and chaotic regions
in the classical phase space. Ideally, in the strict semiclassical limit, $M=-1$ in purely
regular regions, and $M=1$ in pure chaotic regions. 
In practice, we also find various eigenstates (PH functions)
with $-1 < M < 1$, that belong to mixed states.  We studied the joint probability distributions of all combinations of the measures, namely $P(A,R)$, $P(A,M)$ and $P(R,M)$. Our analysis confirms
that $A$ and $R$ are on the average
linearly related. Therefore, the classification of
states and the results on the statistical
properties of the degree of localization for the
chaotic states do not depend
very much on the definition of the localization measure. However, subtle differences allow us to identify interesting mixed-type states and also distinguish between the regular and chaotic states based strictly on comparing the different localization measures, without considering the classical phase portraits. Indeed, measures based on higher order Renyi entropy might also be considered to enhance the differences.

 Very similar localization measures defined on different Hilbert space basis sets have been used to describe multifractality in random matrix models \cite{backer2019multifractal}, many-body localized \cite{mace2019multifractal} and  nonergodic extended states \cite{de2020multifractality}, the transition to chaos in interacting boson systems described by Bose-Hubbard Hamiltonians \cite{pausch2021chaos,pausch2021}, Anderson localization of Rydberg electrons interacting with ground state atoms \cite{eiles2021anderson} to name some examples. Using the various definitions mentioned above, it would be possible to extend our approach and phenomenological descriptions to systems without a clearly defined classical limit.  In particular, we propose to study the relations between the different orders of the localization measures in terms of joint probability distributions, as presented in this paper. The Berry-Robnik picture and the separation of eigenstates into regular and chaotic states is well established for non-interacting low-dimensional quantum systems with a well-defined semiclassical limit, and has been again demonstrated in the present paper. However, even classically, ergodicity and chaos is hard to prove when considering many-body interacting systems, as small barely detectable islands of stability may persist. Naively, one expects that generic many-body interacting systems are ergodic and admit a statistical-mechanical description, yet these types of quantum systems may still display intermediate spectral statistics (between chaos and integrability). Weak ergodicity breaking has also been attributed to so called many-body scarred states \cite{serbyn2021quantum}.  It is not clear whether these states are associated with classical islands of stability, periodic orbit scarring, dynamical localization or some other mechanism. Since clear structures associated with regular and chaotic states are visible in the joint probability distributions of the localization measures in our non-interacting example, it is feasible that some similar structures, if found in the many-body systems, would indicate the existence of states that are associated with islands of stability and would thus elucidate the many-body scarring mechanism. Further interesting research directions are possible in tight-binding billiard models \cite{ulcakar2022}, where it has been shown the tight-binding billiard has similar ergodic properties to its continuous counterpart in the fully-chaotic case.

To conclude, our study of the PH functions confirms that the Berry-Robnik picture of separation into regular and chaotic 
eigenstates is correct. The underlying mechanism is the principle of
uniform semiclassical condensation of the Wigner functions (or
PH functions) \cite{Rob1998} that was developed from Percival's conjecture \cite{percival1973} and Berry's work \cite{berry1977,berry1977semi}. 
The PH functions are asymptotically (in the ultimate semiclassical
limit) either of the regular type or of the chaotic type.
Mixed PH functions at lower energies exist,
and we have studied them in detail.
We have shown that their number monotonically decreases with increasing
energy in the semiclassical limit of high-lying eigenstates.
Moreover, we have quantified this observation by showing
that the relative fraction of mixed-type states decreases as a power law with increasing energy. This is the central result of our paper. Since billiards are a representative example of low-dimensional generic Hamiltonian systems the approach is directly applicable to any quantum mixed-type Hamiltonian systems with a clear classical limit, such as e.g. recently studied Dicke model \cite{WR2020}, the kicked top model \cite{WR2021}, or the three-site Bose-Hubbard model \cite{nakerst2022chaos}.

\section{Acknowledgement}
We thank B. Georgeot, R. Ketzmerick and L. Vidmar for elucidating discussions, T. Prosen for discussions and providing the extensive use of computational facilities and M.T. Eiles for the careful reading of the manuscript.  Č.L. thanks the MPG for its hospitality.
This work was supported by the Slovenian Research Agency (ARRS) under the grant J1-9112.

\bibliography{qChaos.bib}
\FloatBarrier
 
\appendix
\section{Lemon billiard geometry}\label{AppendixA}

\begin{figure}
 \begin{centering}
   \includegraphics[width=5cm]{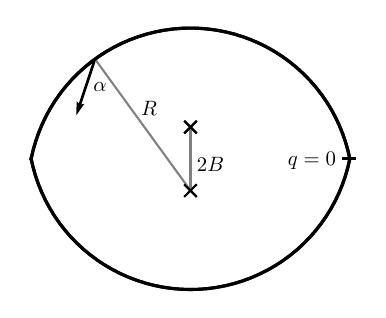}  
   \par\end{centering}
 \caption{Illustration of the lemon billiard geometry. Here $B=0.1953$}
\label{fig: sketch}
\end{figure}
The lemon billiards are defined by the intersection of two circles of equal unit radius with the distance $2B$ between their centers.
The construction is illustrated in Fig. \ref{fig: sketch}. In Cartesian coordinates, the boundary is given by the following implicit equations

\bea   \label{lemonB}
 x^2 + (y+B)^2  = 1, \;\;\; y > 0, \\  \nonumber
 x^2 + (y-B)^2 = 1, \;\;\; y < 0.
\eea
To construct the canonical Poincaré-Birkhoff coordinates $(q,p)$, we take the boundary as the surface of section. The billiard dynamics (a series of line segments linking the collisions) is described by a series of points where the bounce position is given by the arclength $q$ and the corresponding canonical momentum is $p=\sin(\alpha)$. We set the origin of the $q$ coordinate into the right kink (corner) and integrate the length of the boundary counterclockwise up to the collision point. The period $q$ is given by the circumference of the entire billiard boundary

\be \label{perimeter}
{L} = 4 \arctan \sqrt{B^{-2} -1}.
\ee
The area ${A} $ of the billiard is equal to

\bea \label{area}
{A} & = & 2 \arctan \sqrt{B^{-2} -1} - 2B\sqrt{1-B^2}\\ \nonumber
& = & \frac{1}{2} {L} - 2B\sqrt{1-B^2}.
\eea

To compute the fractional measure of the chaotic component, we will first compute
how the area $\chi_{c}$, with one of the methods of Ref. \cite{LozRob2018B}in the two-dimensional
phase space (i.e. the surface of section in the Poincaré-Birkhoff
coordinates) and then calculate the Liouville measure in the full
four-dimensional phase space. The relation between the two is given
by the formula derived by Meyer \cite{meyer1986}
\begin{equation}
\rho_{c}=\frac{\chi_{c}}{\chi_{c}+\left(1-\chi_{c}\right)\kappa},\label{eq:LouvilleMeasure}
\end{equation}
where $\kappa=\frac{\left\langle t\right\rangle _{r}}{\left\langle t\right\rangle _{c}}$
is the ratio between the average return time to the surface of section
(length of the trajectory between collisions divided by the speed
of the particle) on the regular components $\left\langle t\right\rangle _{r}$
and the average of the same quantity on the chaotic component $\left\langle t\right\rangle _{c}$.
For the equivalent formula pertaining to $\rho_{r}=1-\rho_{c}$, we only need
to exchange the indices $r\leftrightarrow c$ and invert the ratio
$\kappa\rightarrow1/\kappa$ in Eq. (\ref{eq:LouvilleMeasure}). In
billiards the surface of section is the billiard boundary and the ratio $\kappa$ is independent of
the speed of the particle. The SOS return time is proportional to
the length of a link of a trajectory between two consecutive collisions.
The ratio $\kappa$ is numerically computed by averaging the length
of a link over a number of collisions and then computing the averages
with regard to the initial conditions.

\section{Recurrence times and stickiness}\label{AppendixB}
Generic mixed-type Hamiltonian commonly exhibit the phenomenon known as stickiness - chaotic orbits intermittently "stick" to islands of stability or other invariant structures for extended periods of time. This results in a slow (sub-exponential) decay of correlations and other observables like recurrence times. Here, we will briefly outline the method for quantifying stickiness used to produce the $S$-plots, based on the statistics of recurrence times. See \cite{Lozej2020,LozejThesis} for a more in-depth explanation of the method and related results on stickiness. We consider the billiard system as a map, thus the time is measured discreetly with the number of bounces (map iterations). Let $\cal A$ be an arbitrary subset of the phase space, for instance a small cell $dq \times dp$. The first recurrence
time to $\cal A$ for a point $a \in \cal A$ is defined as the number of iterations an orbit needs
to return to the same cell for the first time,
\begin{equation}
\tau_{\cal A}=\underset{t>0}{\mathrm{min}}\{t:f^t(a) \in \cal A\}, \label{eq:RecTime}
\end{equation} where $f: (q,p) \rightarrow (q',p')$ is the bounce map.
We are interested in the probability distributions of recurrence times. For chaotic systems, one expects the recurrences to be essentially uncorrelated, and thus the mean recurrence time is the inverse of the area of the test set (Kac's lemma) and the distribution is exponential. If we discretize the phase space into a grid of $N_c$ cells, 
\begin{equation}
P\left(\tau\right)=\frac{1}{N_c}\exp\left(-\frac{\tau}{N_c}\right).\label{eq:ExponentialDist}
\end{equation}
The assumption of completely uncorrelated cell recurrences is a strong one and by definition holds for so-called Bernoulli systems. Chaoticity in the sense of positive Lyapunov exponents is a weaker ergodic property. However, strong empirical evidence suggests the recurrence times generically exhibit an exponential distribution in the bulk of the chaotic component, even in mixed type systems, outside of the sticky areas. This is in agreement with the findings in the so-called random model \cite{robnik1997new}. Stickiness is a consequence of partial transport barriers like cantori. The chaotic orbits become intermittently trapped, and thus short recurrences feature more prominently in the distribution. To quantify this effect, we exploit a special feature of the exponential distribution, namely that its variance is equal to its mean $\sigma=\mu$. The variable $S=\sigma / \mu$ (coefficient of variation) can thus distinguish between exponential and non-exponential distributions of recurrence times. When we consider small areas of the phase space (discretization cells) we may distinguish areas of uniform chaos $S=1$ and sticky areas $S>1$. The easiest way of generating the recurrence time statistics is to just run a sufficiently long chaotic orbit and track the $S$ parameter locally in each cell.

\end{document}